\documentclass[12pt,preprint]{aastex}
\usepackage{graphicx}
\usepackage[latin1]{inputenc}
\usepackage{txfonts}
\usepackage{longtable}
\usepackage{threeparttablex}

\usepackage{float}

\makeatletter
\let\captionbox\@undefined       
\makeatother
\usepackage[justification=centering]{caption}
\usepackage{lscape}
\usepackage{natbib}
\usepackage{array}
\usepackage{gensymb}
\usepackage{hyperref}

\shorttitle{Gausianity of the Velocity Distribution}
\shortauthors{de Carvalho et al.}%%%%% AUTHORS - PLACE YOUR OWN MACROS HERE %%%%%

\begin{document}

\title{Investigating the Relation between Galaxy Properties and the Gaussianity of the Velocity Distribution of Groups and Clusters}

\author{R.R. de Carvalho$^{1}$, A.L.B. Ribeiro$^{2}$, D. Stalder$^{3}$, R.R. Rosa$^{3}$, A.P. Costa$^{2}$,  T. Moura$^{1}$}
  
\affil{$^1$INPE/MCT, Divis$\tilde{a}$o de Astrof\'isica, S. J. dos Campos, Brazil}

\affil{$^2$ UESC, Laborat\'orio de Astrof\'isica Te\'orica e Observacional, Ilh\'eus, Brazil}
  
\affil{$^3$INPE/MCT, Laborat\'orio Associado de Computa\c c$\tilde{a}$o e Matem\'atica Aplicada, S. J. dos Campos, Brazil}

\begin{abstract}

We investigate the dependence of stellar population properties of galaxies on group dynamical stage for a subsample of Yang catalog. We classify groups according to their galaxy velocity distribution into Gaussian (G) and Non-Gaussian (NG). Using two totally independent approaches we have shown that our measurement of Gaussianity is robust and reliable. Our sample covers Yang's groups in the redshift range 0.03 $\leq$ z $\leq$ 0.1 having mass $\geq$ 10$^{14} \rm M_{\odot}$. The new method, Hellinger Distance (HD), to determine whether a group has a velocity distribution Gaussian or Non-Gaussian is very effective in distinguishing between the two families. NG groups present halo masses higher than the G ones, confirming previous findings. Examining the Skewness and Kurtosis of the velocity distribution of G and NG groups, we find that faint galaxies in NG groups are mainly infalling for the first time into the groups. We show that considering only faint galaxies in the outskirts, those in NG groups are older and more metal rich than the ones in G groups.  Also, examining the Projected Phase Space of cluster galaxies we see that bright and faint galactic systems in G groups are in dynamical equilibrium which does not seem to be the case in NG groups. These findings suggest that NG systems have a higher infall rate, assembling more galaxies which experienced preprocessing before entering the group.

\end{abstract}

\section{Introduction}

From the time cosmologists identified galaxies as separate units in the Universe, they started asking how, where, and when they form. In the last decades, significant strides have been made in identifying the factors which establish galaxy's morphology and how star formation proceeds since early times. Initial conditions in the Universe set the way galaxies formed and evolved and today semi analytical models, SAM \citep[e.g.][]{WHFR91,COLE94,CAT07} and hydrodynamical simulations \citep[e.g.][]{RY93,SWTK01} help us interpret the data we have been gathering in the recent past. Another way of studying how galaxies evolve is by measuring their properties at different redshifts and environments and trying to match them to the best models.

Early and late type galaxies are located preferentially in opposite environments, a fact described by the morphology-density relation  \citep{Oemler,DRESS80}. At first sight, it implies that internal properties of galaxies are modified by the environment (the ``nature'' versus ``nurture'' debate). Field galaxies would exhibit characteristics set as they were born while in denser systems (groups and clusters) processes like ram-pressure, starvation and harassment would transform the system. Over the last two decades, observations have shown that star formation is enhanced already in the infall regions of clusters {\it wrt} the field, exhibiting the role of the environment \citep[e.g.][]{KAU04,vdW10,MMR11,WTC13}. These investigations show that the fraction of quiescent galaxies varies significantly with the environment, namely higher in clusters than in low density groups \citep[e.g.][]{BALOGH2004}. However, galaxy properties (e.g. morphology, color) also seem to be more strongly related to stellar mass \citep[e.g.][]{BALOGH2009,Peng2010,Woo13}, recovering the idea that nature is the key factor in determining the way galaxies evolve. But galaxy stellar mass correlates with environment - more massive galaxies are more likely to be found in high-density regions. Therefore, it seems impracticable to distinguish the effects of  ``nature'' from those of ``nurture''.

Another important piece of information about galaxy evolution comes from the fact that the fraction of blue galaxies (measured within a radius containing 30\% of the projected galaxy distribution) in clusters increases with redshift  \cite{BUOE78}, the so called BO effect. This result was later confirmed by \cite{Margo2001} and \cite{KOBO01} determined that the relation extends up to z$\sim$1. The BO effect might be seen as consequence of the increase of the cosmic star formation rate (SFR) up to z = 1 \cite[e.g.][]{Madau96}, namely, increasing fraction of blue galaxies in clusters in the redshift range of $0<z<1$. However, \cite{ELLIN01}, examining clusters between $0.18<z<0.55$, find that the fraction of blue galaxies within half of the virial radius from the center of the cluster does not change with redshift, implying that the BO effect is not determined by galaxies in the cluster core. More likely, we are seeing blue galaxies falling in from the very low density regions and the higher fraction of blue galaxies implies larger infall rate onto the cluster. Thus, it is clear that the environment is responsible for part of the way galaxies look like today.

The task of defining environment is intimately associated to the definition of equilibrium state of a gravitational system, which in turn is described by a Maxwell-Boltzmann distribution function \citep[e.g.][]{Ogorodnikov,Lynden-Bell67}. In phase-space coordinates this translates into a gaussian function. N-body numerical experiments \citep{MeHe2003,HAN05} also support this conclusion. From the observational viewpoint, it is extremely difficult to determine when a velocity distribution differs from normality \citep[e.g.][]{Beers:90}, especially for the low-multiplicity systems. \cite{HOU09} considered three figures of merit (Anderson-Darling, Kolmogorov-Smirnov and $\chi^{2}$-test) aiming to find which statistical tool distinguishes better between gaussian and non-gaussian groups. Using Monte Carlo simulations and a sample of groups selected from CNOC2 \citep[Canadian Network for Observational Cosmology, ][]{LinEtal1999}, they found the Anderson-Darling test to be much more reliable at detecting real departures from normality. Also, gaussian and non-gaussian groups exhibit distinct velocity dispersion profiles, suggesting different dynamical stages. About 68\% of the CNOC2 groups are found to be gaussians. Hence, the choice of the statistical test to be applied on data is crucially important in the subsequent analysis of galaxy groups. It is important to keep in mind that sample size is a potential problem for all the hypothesis tests presented in the literature.

Usually, in most works, environment is mainly characterized by galactic density. However, more recently several investigations have discussed the importance of establishing the dynamical state of a group/cluster  \citep[e.g.][]{MMR11,Einasto12b,Einasto12a} using the velocity distribution which may qualify better for a robust descriptor of environment. \cite{Einasto12a} examining a sample of rich clusters selected from SDSS-DR8 (Sloan Digital Sky Survey - 8$^{th}$ Data Release), using a FoF algorithm (Friends of Friends), find that most clusters are dynamically young based on their amount of substructure, large peculiar velocities, and non-gaussianity of their velocity distributions, emphasizing that the halo model (which assumes virialization) does not explain the cluster properties.This result is reinforced by the work of \cite{Maccio07}. Considering the importance of establishing the gaussianity of the velocity distribution of galaxies in clusters, \cite{Ribeiro2013} propose a new definition of gaussianity of the velocity distribution (Hellinger Distance), based on the distance between empirical and theoretical distributions. They find that in gaussian groups, there is a significant difference between the galaxy properties of the inner and outer galaxy populations, suggesting that the environment is actively affecting the galaxies \citep[see also][]{RobertsParker2017}. On the other hand, in non-gaussian groups there is no segregation between the properties of galaxies in the inner and outer regions, which might indicate that the properties of these galaxies still reflect primordial physical processes prevailing in the environment. \cite{CohenEtal2014} found that the fraction of star-forming galaxies is higher in multi-component (non-gaussian) clusters when compared to the one-component systems (gaussian), and increases with clustercentric distance. Later, \cite{CohenEtal2017}, examine the relation between star formation, substructure and supercluster environment and find that cluster star formation is mainly determined by not only the dynamical youth of the cluster (younger systems display higher star formation) but also by the supercluster environment (star formation is lower in the supercluster core). Although much have been done in recent years in this area, the relation between the gaussianity of the velocity distribution of a galactic system and the internal properties of the member galaxies is still unclear. In this work, we examine this relation in detail considering the new HD parameter presented by \cite{Ribeiro2013}. 

This  manuscript  is  organized  as follows: Section 2 defines the sample and  present the data characterizing the galaxy properties; Section 3 discusses how we measure gaussianity and its reliability. Section 4 presents the study of the groups of the Yang sample. In Section 5 we discuss the implications of the gaussianity measurement introduced here to the main question of how galaxies are affected by the environment and Section 6 summarizes the principal findings of this investigation. Throughout the paper, we adopt the $\Lambda$CDM cosmology with H$_{\rm o}$ = 72 Km s$^{-1}$ Mpc$^{-1}$, $\Omega_{\rm M}=0.27$, $\Omega_{\Lambda}=0.73$.

\section{Sample and Data\label{SecSampleAndData}}

The velocity distribution of a galactic system carries important information about its dynamical state. However, the complexity of the large scale structure and the difficulty in defining unbiased samples limit our understanding of the interplay between the process of virialization of a system and the properties of galaxies bounded to it. In this work, we focus our attention on the question of how the gaussianity (in a state of dynamical equilibrium) of the velocity distribution in a group/cluster is connected to the properties of the member galaxies. To study the updated group catalog of  \cite{Yang07} (hereafter Y07), we selected galaxies from SDSS-DR7 with 0.03 $\leq$ z $\leq$ 0.1 and $r$ magnitudes brighter than 17.78, which is the spectroscopic completeness limit of the survey, guaranteeing that we probe the luminosity function up to M$^{*}+1$ for all systems. The lower limit in redshift is imposed to avoid aperture effects in the stellar population parameters measured within a fixed aperture of 3 arc sec (diameter) used in the SDSS.  In the analysis that follows we consider two specific luminosity domains: Bright means $\rm M_{\rm r} \leq -20.55$, which is the limiting absolute magnitude corresponding to the spectroscopic completeness of SDSS-DR7 at z = 0.1, namely the bright regime probes the systems up to $\rm M^{\star}+1$ (Blanton et al. 2006); Faint means $-20.55 < \rm M_{\rm r} \leq -18.40$, where the limiting absolute magnitude corresponds to the spectroscopic completeness of SDSS-DR7 at z = 0.04. Thus, the faint regime is analyzed only for systems in the $0.03 \leq \rm z \leq 0.04$ domain and it probes the luminosity function down to $\sim \rm M^{\star}+3$.

The parameters characterizing the stellar populations were obtained by running the spectral fitting code starlight \citep{CMSS05} on 570,685 galaxies for which zWarning=0 in the SDSS DR7 database. We derived ages (of stellar populations of galaxies), metallicities, internal extinction and stellar masses, after the observed spectra are corrected for foreground extinction and de-redshifted, and the single stellar population (SSP) models are degraded to match the wavelength-dependent resolution of the SDSS spectra, following prescription in \cite{Spider2010}. We adopted \cite{Cardelli1989} extinction law, assuming R$_{\rm V}$ = 3.1. We used SSP models based on the Medium resolution INT Library of Empirical Spectra \citep[MILES - ][]{SancherBlazquez2006}, using the code presented in \cite{Vazdekis2010}, using version 9.1 described in \cite{FalconBarroso2011}. They have a spectral resolution of $\sim$2.5 \AA, nearly constant with wavelength. Models were computed with \cite{Kroupa2001} Universal IMF with slope = 1.30, and isochrones by \cite{Girardietal2000}. The basis grids cover ages of 0.07 to 14.2 Gyr, with constant log(Age) steps of 0.2. We selected SSPs with metallicities [Z/H] ={-1.71,-0.71,-0.38,0.00,+0.20}. The stellar masses are computed within the fiber aperture and extrapolated to the full extent of the galaxy by computing the difference between fiber and model magnitudes in the z band. Then we have log(M$_{*}$) = log(M$_{*}$)$^{\prime}$ + 0.4 (m$_{fiber,z}$ - m$_{model,z}$).

Considering the importance of these stellar population parameters for the analysis presented in this paper, we assessed the uncertainty in Age, [Z/H], and stellar mass in two different ways. First we compare our estimates to those available in SDSS-DR12 \footnote[1] {skyserver.sdss.org/dr12/en/help/browser/browser.aspx\#\&\&history=description+stellarMassPCAWiscBC03+U}. The method used to obtain the three stellar population parameters used here is described in \cite{ChenEtal2012} and is based on PCA (Principal Component Analysis) and using the stellar population synthesis models of \citep[BC03 - ][]{BruzualCharlot2003} to generate a library of model spectra. Direct comparison gives the following residuals: $\Delta$ Age (Our - Their) = 3.46$\pm$2.74 Gyr; $\Delta$ [Z/H] = -0.006$\pm$0.082; and a difference of 0.10$\pm$0.081 dex in log of stellar mass. These differences do not seem significant except for Age, which is a longstanding problem found by other authors when comparing MILES and BC03 \citep[][]{KolevaEtal2008} and is still not clear what is the source of disagreement. A second way of comparing and actually getting a more robust uncertainty estimate is done by using repeated observations of the same galaxy. We queried SDSS-DR7, the source of data for this work, and searched for galaxies with two or more observations in the same redshift and magnitude range of the primary galaxy sample used here requiring a  S/N of the spectra to be $>$ 20. This results in 6148 repeated observations of 2543 galaxies. The residual distributions indicate the following uncertainties: : $\Delta$ Age (Our - Their) = -0.01$\pm$1.19 Gyr; $\Delta$ [Z/H] = 0.0038$\pm$0.0439; and a difference of -0.001$\pm$0.0746 dex in log of stellar mass.  These uncertainties are consistent with those obtained comparing with SDSS-DR12, except for Age, which is not surprising considering that BC03 yields younger ages in comparison with MILES \citep[see][]{KolevaEtal2008}.

The dynamical analysis of each group in Yang sample was done using the shift-gapper technique as in \cite{Lopes2009}. Here we briefly describe how this technique works. The first step is to use the center (right ascension and declination) and redshift of the group from Yang catalog. However, this is used only to select galaxies around the cluster (querying SDSS-DR7) and feed the shift gapper technique. One of the main advantages of the shift gapper technique is that no hypotheses about the dynamical status of the system is made \footnote[2]{{Also, using shift gapper we avoid undesirable conceptual problems introduced by the FoF algorithm like the indeterminacy of choosing single-linkage, average-linkage or complete-linkage \citep[see][]{EverittEtal2011}}}. The algorithm follows the one in Fadda et al. (1996). We apply the Gapper technique in radial bins with sizes of 0.42h$^{-1}$Mpc or larger, guaranteeing at least 15 galaxies per bin. The procedure is reiterated until no more interlopers are found and the final list of members is used to estimate cluster properties \citep[see][]{Lopes2009} like velocity dispersion, radius (R$_{200}$) and virial mass (M$_{200}$). The analysis is done within a maximum distance of 2.5h$^{-1}$Mpc (3.47 Mpc for h = 0.72) from the cluster center. Our shift gapper code has been compared to a set of 24 galaxy-based cluster mass estimation techniques and proved to be among the best three \citep[][]{Oldetal2015}. From this analysis we find that the error associated to our mass estimate, M$_{200}$,  is $\sim$ 0.22 dex. 

\begin{figure}[!h]
	\begin{center}
		\includegraphics[width=0.7\textwidth]{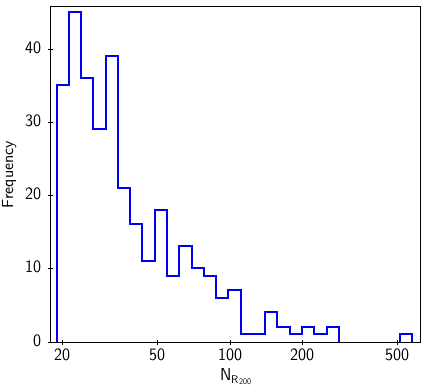}
		\caption{Distribution of richness, as measured by the number of galaxies within R$_{200}$ down to the limiting magnitude probed in this study.\label{fig1}}
	\end{center}
	
\end{figure}

We tested membership against which cluster center to use. The difference in number of members per group when using either Yang's original center or the one re-estimated by the shift gapper technique is in average 3 galaxies. This is important to quantify what is the impact of the center determination on the gaussianity of the velocity distribution and on the distribution of galaxies in the projected phase space. In the analysis that follow we use the shift gapper center. Only systems richer than 20 galaxies (within R$_{200}$) are used in this work (see Section 3.1 for more details on why we chose this lower limit). Considering these constrain in redshift (0.03 $\leq$ z $\leq$ 0.1) and richness, we end up with 319 groups. In Table \ref {tab:long} we present all parameters describing the sample used in this work. Column (1) lists the identification number in Yang catalog; Columns (2) and (3) list right ascension and declination in J2000; Column (4) lists the mean redshift of the group as measured with the shift gapper technique; in Columns (5) and (6) we present the dynamical mass and the virial radius as obtained through the shift gapper algorithm and dynamical analysis; Column (7) lists the number of member galaxies within R$_{200}$ down to the limiting magnitude of the catalog, N$_{\rm R_{200}}$ ; Finally, in Columns (8) and (9) we show which group has X-ray counterpart in BAX (http://bax.ast.obs-mip.fr/) or NORAS (Northern ROSAT All-Sky), \citep{Bohringeretal2000}, and REFLEX (ROSAT-ESO Flux-limited X-ray), \citep{Bohringeretal2004}. More details on the X-ray association will be presented in Section 4. In Fig ~\ref{fig1}, we present the distribution of richness for the groups studied in this work, N$_{\rm R_{200}}$.

\section{Characterizing the velocity distribution of galaxies in Groups/Clusters}

The large scale structure of the Universe exhibits clustering covering the whole mass domain. Also, the morphology-density relation and the BO effect indicate that structural parameters and stellar populations of galaxies may vary according to the environment where these systems are located. Throughout the literature, environment is mostly intuitively associated to local density, although this may not be effective in characterizing the role of it on the evolution of a galaxy. It is important to bear in mind that groups/clusters are not isolated entities; massive clusters, for instance, are seeing in cosmological simulations as intersections of filaments. Therefore, it is expected that these systems are always accreting small galaxies (or groups), which may modify the underlying velocity distribution. It is quite likely that these accretions: 1) alter the dynamics of the system; 2) change the properties of the galaxies which were already in the group/cluster; and 3) bring new galaxies that may have structure and stellar content significantly different from the ones formed {\it in situ}. This complexity is modulated with the physical mechanisms operating in clusters of different masses and different stages of dynamical evolution, like ram-pressure, starvation and harassment. The fundamental question here is: Is there a relation between galaxy properties and deviations from gaussianity of the velocity distribution of the galaxies in galactic systems ?

In a previous work \cite{Ribeiro2013}, we introduced a new estimator of the distance between the empirical velocity distribution of galaxies in a group and the theoretically expected Gaussian distribution function, the so called Hellinger distance - a stable approximation to the Fisher information metric \cite[e.g.][]{Amari85}. We find that in gaussian groups, there is a significant difference between the galaxy properties of the inner and outer galaxy populations, suggesting that the environment is actively affecting the galaxy properties. Also, in non-gaussian groups there is no segregation between the properties of galaxies in the inner and outer regions. Recent works show that multimodal velocity distributions may be very common in galaxy systems \cite[e.g.][]{Ribeiro2011,HOU12,Einasto12a}. However, multimodality depends on the separation and widths of the modes  \cite[see][]{AsBiZe94}; thus, it is of paramount importance to assess the statistical reliability in detecting modes in a velocity distribution to conduct a comparative study of how galaxy properties depend on the characteristics of the velocity distribution.

\subsection {How to Reliably Detect a Non-Gaussianity in Velocity Distributions ? \label{SecReliability}} 

In this investigation, we assume that bimodal expression patterns may result from: two big groups interacting; a big group accreting a small one; or may be a perturbation of a single gaussian distribution. Unimodal distributions would indicate closeness to virialization. The problem of finding multiple modes (gaussians, for simplicity) in a distribution is a longstanding one. \cite{HEL1904}, considers the mixture of two normal distributions, with means $\mu_{1}$ and $\mu_{2}$, and common variance $\sigma$, and proves that the mixture will be seen as unimodal ${if~ and~ only~ if}$ $\mid\mu_{1}-\mu_{2}\mid < 2\sigma$. This result is not generalized for the case where the two modes have different variances \cite[e.g.][]{SCH02}. 

An important point to consider when examining a velocity distribution is that we can either try to identify multiple modes (gaussians), which mixture justifies the distribution or we can directly measure how far from a gaussian the distribution is. In the following, we investigate these two approaches using two specific techniques by creating realizations which are perfect gaussian mixtures. Although this simplifying assumption may not represent what we observe in real clusters, it serves as a guidance for how these methodologies respond to typical values of the parameters involved in the multimodality modeling. 

\subsubsection{MCLUST\label{secMclust}}

{MCLUST is a powerful R package for modeling data as a Gaussian finite mixture \citep[][]{FraleyAndRaftery2002}. In the model-based approach to clustering, each component of a finite mixture density is usually associated with a group or cluster. Most applications assume that all component densities arise from the same parametric distribution family, although this need not be the case in general. A common used approach is the Gaussian mixture model (GMM), which assumes a (multivariate) Gaussian distribution for each component. GMM is a better version of KMM (Kayes Mixture Model), a mixture modeling code for detecting bimodality in astronomical applications \citep[][]{AsBiZe94}. The KMM algorithm assumes that an input sample is described by a sum of two Gaussian modes and calculates the likelihood of a given data point belonging to either of the two modes.  It also calculates the likelihood ratio test as an estimate of the improvement in going from one Gaussian to two Gaussian distributions. A different test of uni-modality was proposed by \cite{Hartigan1985}  and first used in astronomy by \cite{Gebhardt1999}. It is called dip statistics and what it does is to determine the maximum distance between the cumulative input distribution and the best-fitting unimodal distribution. This test is similar to KS but it searches specifically for a flat step in the cumulative distribution function, which corresponds to a ÒdipÓ in the histogram representation. Both methods have advantages and disadvantages. In the present work, we decided to use MCLUST to probe multimodality after the work of \cite{Ribeiro2013} who verified that MCLUST is a better choice to study samples with sizes $\geq 20$. Reinforcing this option, \cite{Muratov2010}  found that the dip test appears less powerful than the GMM algorithm when modeling the metallicity distribution of globular clusters in the Galaxy. A short explanation of how MCLUST works is presented below.}

A given velocity distribution $\bf{v}=(v_1,....,v_n)$ can be seen as a random sample of a univariate random variable $V$ whose density function is expressed as a mixture of gaussians:

\begin{equation}
p(x_i\mid \theta) = \sum^{N_M}_{k=1} \pi_{k} G (x_i\mid \mu_k,\sigma_k)
\end{equation}

\noindent where $\pi_k$ is the proportion of samples in the groups, ($\mu_k,\sigma_k$) are the mean and standard deviation of the gaussian k and $\theta$ denotes the set of all parameters. The number of modes can be inferred by the EM (Expectation Maximization) algorithm to learn the parameters for a certain range of different ${N_M}$ (number of normal modes). Although most algorithms for fitting mixtures (where we do not know the number of components) use EM, certain issues are present: 1) EM strongly depends on initialization - this is usually fixed by using multiple random starts and choosing the highest likelihood solution \cite[e.g.][]{McLachlanPeel2000}; and 2) EM sometimes converges to the boundary of the parameter space - this problem is usually solved by the use of soft constraints on the covariance matrices \cite[e.g.][]{Kloppenburg1997}. In our case, since we do not expect to have too many groups with a large number of modes, this is not a critical issue. {The optimal ${N_M}$ (model selection) is estimated using the Bayesian Information Criterion (BIC) score \citep[][]{Yeung2001, KassAndRaftery1995}. MCLUST outputs $\mu_{k}$, $\sigma_{k}$ and $\pi_{k}$, for $k$ running from 1 to $N_M$. We also define the distance between the first two most dominant modes as $\delta = \mid \mu_{1}/\sigma_{1} -  \mu_{2}/\sigma_{2}\mid$. 

{MCLUST has made its entrance in astronomy with the paper by \cite{Einasto10}}, but in other fields is already very popular, especially biology}. For instance, \cite{Wang2009} uses MCLUST to identify genes with bimodal expression patterns and in order to do this they run a series of simulations to understand the limits of applicability of the method. First, they generate unimodal distributions with $n$ points (from 50 to 300) and conclude that MCLUST, as well MCMC (Markov Chain Monte Carlo), yield very low false positive rate, $<$$ 3\%$ (type I errors, those that occur when the null hypothesis is true but rejected). This means that running MCLUST on samples with more than 50 points results in detecting unimodal distributions with high statistical significance. Second, they determine how reliable their approach is when dealing with truly bimodal simulated measurements. In this case, $\delta$ and $\pi$ are key factors establishing the performance of the method as well as $n$. For $30\%\leq\pi\leq 70\%$, when $\delta \ge 4$ MCLUST correctly identifies bimodal distribution 98\% of the times, namely a low false negative rate (type II errors, those that occur when the null hypothesis is false and erroneously taken as true). For $10\%\leq\pi\leq 90\%$ and $\delta \ge 4$ MCLUST drops to 83\%. These results are very intuitive - even if two modes are very separate (large $\delta$), a very small $\pi$ would indicate that the smaller mode becomes statistically non-significant diminishing our ability to detect a true bimodal distribution. \cite{Wang2009} conclude that for $\pi\leq 0.1$ or $\pi\geq 0.9$ and a small sample size ($\leq$100 points), the false negative rate will be large even for large $\delta$. 

\begin{figure}[!h]
%	\begin{center}
		\includegraphics[width=1.0\textwidth]{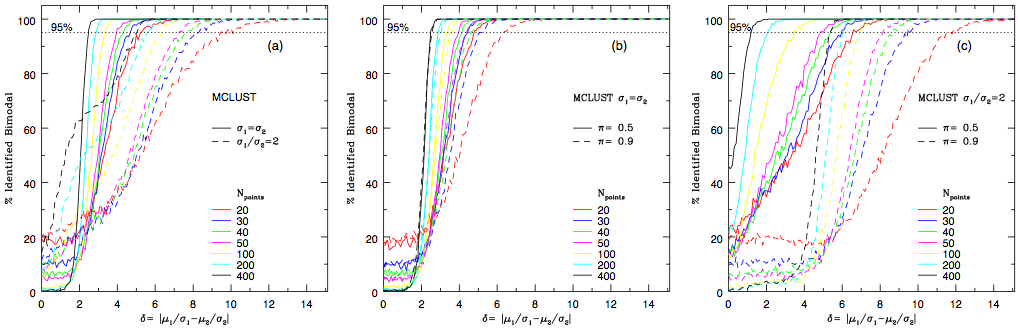}
		\caption{Performance of MCLUST in simulated bimodal data set and its dependence on different sample size, N$_{points}$, proportion in one group, $\pi$, varying from 0.5 to 0.9, in steps of 0.1, and the FWHM (or $\sigma$) of the gaussian. We display in all three panels the percentage of identified bimodal distributions as a function of $\delta$. A dotted line is displayed in each panel identifying the 95\% recovery. Different colors for different dashed and solid lines indicate different N$_{points}$ and different conditions for $\sigma$ and $\pi$. \label{fig2}}
%	\end{center}
	
\end{figure}

Here, we repeated Wang's experiment by testing how reliable MCLUST is in recovering bimodal distributions. For a given total number of points (N$_{\rm points}$) defining both gaussians, a given ratio of $\sigma's$ and a given separation between the gaussians (expressed by $\delta$, as defined above) we created 1000 realizations with $50\%\leq\pi\leq 90\%$, with 200 realizations for each value of $\pi$. This domain in $\pi$ was used due to its symmetry nature. The result of this experiment is show in Fig ~\ref{fig2}a, where we can see that is far easier to detect bimodal distributions with similar $\sigma's$, regardless the number points defining the whole distribution. We also confirm the fact that for $\delta \le 2$ the ability of MCLUST in recovering bimodal distributions drops significantly in all cases. In conclusion, MCLUST depends on N$_{\rm points}$, $\delta$ and the ratio $\sigma_{1}/\sigma_{2}$. To measure how sensitive MCLUST  is to $\pi$, we run another two experiments, where we fix the ratio $\sigma_{1}/\sigma_{2}$ and create again 1000 realizations, but this time with a fixed value of $\pi$, 0.5 and 0.9, extreme cases of the proportion in one group. As it is clearly seen from Figures \ref{fig2}b and \ref{fig2}c, MCLUST performs better when $\sigma's$ are similar. These results indicate that the final reliability of MCLUST in finding bimodal distributions depend on all different parameters, some of them more important than others. We will return to this point in Section \ref{comparingMclusttoHD}.

\subsubsection {Hellinger Distance\label{secHD}}

The Hellinger Distance (HD) was first introduced in astronomy by \cite{Ribeiro2013}, studying the degree of gaussianity of the velocity distribution of galaxies in groups. The idea behind the HD parameter is as follows. Consider ($\Omega$,B,$\nu$) to be a measure space \cite{halmos1950measure}, and P the set of all probability measures on B, assumed continuous with respect to $\nu$. For two probability measures P$_{1}$, P$_{2}$ $\in$ P, the Bhattacharyya \footnote[3]{An Indian statistician who worked in the 1930s  at the Indian Statistical Institute} coefficient between P$_{1}$ and P$_{2}$, measuring the closeness of two probability distributions, is defined as:

\begin{equation}
p(P_{1},P_{2}) = \int_{\Omega} {\sqrt{{\frac{dP_1}{d\nu}}\cdot{\frac{dP_2}{d\nu}}}d\nu}
\end{equation}

\noindent The HD is then derived using the Bhattacharyya coefficient. For two discrete probability 
measures P and Q, with densities {\sl p} and {\sl q} we can write HD as

\begin{equation}
HD^2 (p,q) = 2 \sum_{x} \Big[\sqrt{p(x)} - \sqrt{q(x)}\Big]^2
\end{equation}

\noindent where $x$ is a random variable. The HD satisfies the inequality 0 $\leqslant$ HD $\leqslant$ $\sqrt{2}$ but some authors prefer to normalize the range (e.g. LeCam 1986). We estimate HD using codes available in R environment under the library distrEx \citep{Ruckdeschel2006}.

For two continuous analytic functions, estimating HD is straightforward from equation 2. However, to compute HD between (empirical) data and a continuous distribution, an appropriate calibration of the metric is required. The R code to estimate HD smooths the input observed distribution using a kernel of size equal to $\sigma_{r}/2$, where $\sigma_{r}$ is a robust estimate of the standard deviation of the distribution (the factor 2 was determined empirically). Calibration in this context means establishing the locus separating G from NG and measuring how HD depends on the number of points representing the distribution. Here, we proceed in the following way: 1) for a given number of points, N, we create 1000 realizations of a gaussian distribution with $\mu$ = 0 e $\sigma$ = 1. Figure \ref{fig3} shows how HD varies with the number of points defining the gaussian distribution. As we can see, the median HD, computed from the 1000 realizations, decreases with N (green line). As N goes to infinity HD goes to 0 since at this limit we would be measuring the distance between two perfect gaussians, which by construction is 0; 2) also, for a given N we determine the threshold between G and NG as the median$+ 3\sigma_{HD}$, where $\sigma_{HD}$ is computed from the quartiles of the distribution of HD for a given N (red line). This is our final rule to establish when a given observed or simulated dataset is G or NG. An important caveat is that the input distribution has to be normalized ($\mu$ = 0 e $\sigma$ = 1) for internal consistency in the R code measuring HD.

We took the same set of realizations used to study the performance of MCLUST and measured how HD is able to distinguish G from NG simulated distributions and how sensitive this method is to $\pi, \delta, \sigma_{1}/\sigma_{2}, \rm and ~ N_{\rm points}$. Figures \ref{fig4}a, b and c show the results in the same way as presented for MCLUST. We can see that HD has the same dependence on all different parameters as MCLUST.

\subsubsection {Comparing MCLUST to Hellinger Distance\label{comparingMclusttoHD}}

We chose MCLUST and HD first of all because they represent robust statistical approaches already used in other branches of science and there are sufficiently stable algorithms written for them. Also, they are two totally distinct approaches to identify bimodality (non-gaussianity). Table \ref{TablecomparisonMCLUSTHD} summarizes what is shown in Figures \ref{fig2} and \ref{fig4}. The performance here is measured by the value of $\delta$ when the percentage of identified bimodal distributions is 95\%, namely the ability of a given method to detect two gaussians as they approach each other. The general behavior in both cases is that as N$_{\rm points}$ gets larger both methods can distinguish two gaussians at smaller $\delta$ regardless of $\pi$ and $\sigma's$.  For $\pi$ ranging from 0.5 to 0.9, HD performs slightly better than MCLUST, independent of the $\sigma's$. The same behavior holds true when we fix $\pi = 0.5$, which is the best possible proportion of number of points in both gaussians. For $\pi = 0.9$, which is a limiting case when one gaussian dominates the other (the worst proportion), HD and MCLUST are very similar in detecting bimodality. In summary, although based on idealized realizations, these results show that in the extreme cases ($\pi = 0.5$ and $\pi = 0.9$) HD and MCLUST perform similarly and for $0.5 \leq \pi \leq 0.9$ HD performs better specially when N$_{\rm points}$ is large and $\sigma's$ are different.

\subsubsection {How reliable is the measurement of gaussianity ?\label{secDisMclustHD}}

The results presented in the previous section are based on idealized distributions where bimodality is
defined by the sum of pure gaussian distributions. However, when examining real distributions of line of sight (hereafter LOS) peculiar velocities of galaxies in clusters we do not have any $a ~priori$ information on the underlying distribution. Thus, it is of paramount importance to establish the variance of the measured gaussianity based on the observed data.

To estimate how our measurement of Gaussianity may vary, we adopt a bootstrapping approach, where we randomly draw from the LOS peculiar velocity distribution the same number of data points but with replacement, and run MCLUST and HD in the same way as described previously. For each group, this process is repeated 1,000 times and each time we ask whether the distribution is G or NG. In the case of HD, the answer is straightforward and the system is G or NG depending on the percentage of which is larger than 50\%. As for MCLUST, G is when the number of gaussian modes found is one, otherwise is NG. The important aspect of this approach is that in the end we set the distribution as G or NG with an associated probability, which later will be used as a weight when we examine the properties of galaxies in G and NG systems.

\section{Studying the Yang's Group Catalog}

\begin{figure}[!h]
	\begin{center}
		\includegraphics[scale=0.70]{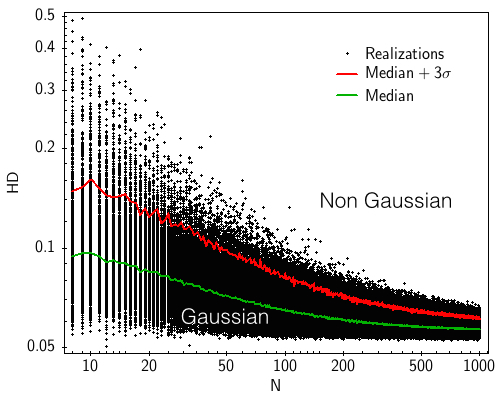}
		\caption{Calibration of the relation between HD and the number of points, N,  sampling the distribution. Green solid line indicates the median of HD for a given N. Red solid line displays the median of HD + 3 $\sigma$, where $\sigma$ is measured for a given N. This is the line used to separate G from NG systems.
		.\label{fig4}}
		
	\end{center}
	
\end{figure}

\begin{figure}[!h]
	\begin{center}
		\includegraphics[width=1.0\textwidth]{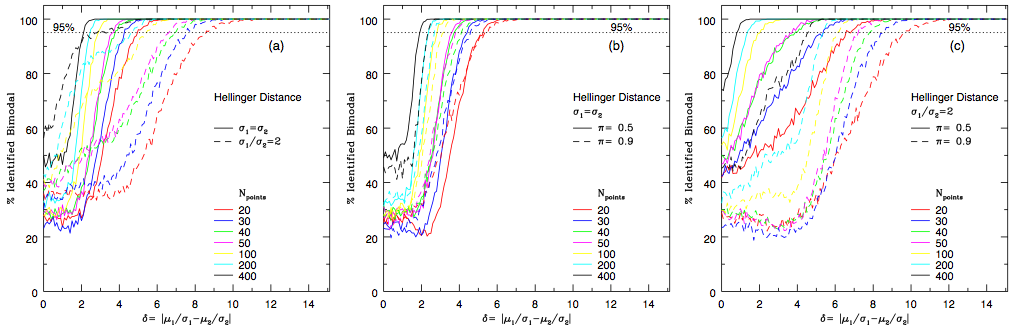}
		\caption{The same as in Figure \ref{fig2} but for the HD measurement of Gaussianity.\label{fig3}}
	\end{center}
	
\end{figure}

We use the techniques described earlier to study the dynamical state of the groups/clusters presented in the updated catalog of galaxy groups of  Y07 by measuring the gaussianity of their LOS velocity distribution. More specifically, the group catalog is based on a sample of 593736 galaxies with available redshifts from SDSS-DR7, supplemented with additional 3115 galaxies with redshifts from different sources. Although this catalog provides mass estimates for all groups, the only information we used was position on the sky and mean redshift. As described in Section \ref{SecSampleAndData}, we use shift-gapper technique to reevaluate the dynamical mass of the groups, their virial radius and membership. We study the velocity distribution of only groups with at least twenty members within R$_{200}$, which means 319 systems. As we can see from Figures \ref{fig2} and \ref{fig4} even for systems with twenty galaxies we expect to detect gaussianity with high statistical significance as long as they  are bimodal with $\delta\geq$4, and $\sigma_{1}$ similar to $\sigma_{2}$, regardless if we use MCLUST or HD. It is important to note that the estimations presented in previous section should be seen as expectations since real distributions can be very different from idealized gaussian distributions, ultimately affecting our ability to detect non-gaussianity which is critically dependent on  the $\sigma_{1}/\sigma_{2}$ ratio and on $\pi$. 

We investigate how MCLUST and HD perform when applied to Yang's catalog of groups as a function of $\delta$. Considering all 319 systems, $\delta$, as measured by MCLUST, varies from 0 to 4.9. But as we learned from Section \ref{secMclust}, when $\delta$ gets smaller than 2 the reliability of distinguishing bimodal distributions drops very fast (See Figures \ref{fig2} and \ref{fig4}), except when we have a large number of galaxies in the system (See Table \ref{TablecomparisonMCLUSTHD}). Thus, considering all 319 groups, the agreement between MCLUST and HD is 66\%, due to the inaccuracy of both methods to detect small deviations of gaussianity, although HD performs better than MCLUST specially for larger N$_{\rm points}$. For $\delta\ge 1.3 $ (27 groups) the agreement is 75\% and if we require an agreement of 90\% only 10 systems are left with $\delta\ge 1.7$. As we discussed in Section \ref{comparingMclusttoHD}, MCLUST is more stringent as it tries to identify multiple gaussians in the distribution, while HD measures deviations from gaussianity. From Table \ref{TablecomparisonMCLUSTHD}, we notice that using HD we reach a certain reliability at a smaller $\delta$ regardless of the pair ($\pi$,$\sigma_{1}/\sigma_{2}$). Therefore, we decided to use HD from now on as the measure of gaussianity of the LOS velocity distribution of the Yang groups. As mentioned before in Section \ref{SecSampleAndData}, changing the center of the cluster results in a small difference in the number of members per group. We tested how that impacts on the gaussianity measurement and found that not a single group changed its HD or MCLUST assignment. 

\begin{figure}[!ht]
	\begin{center}
		\includegraphics[scale=0.7]{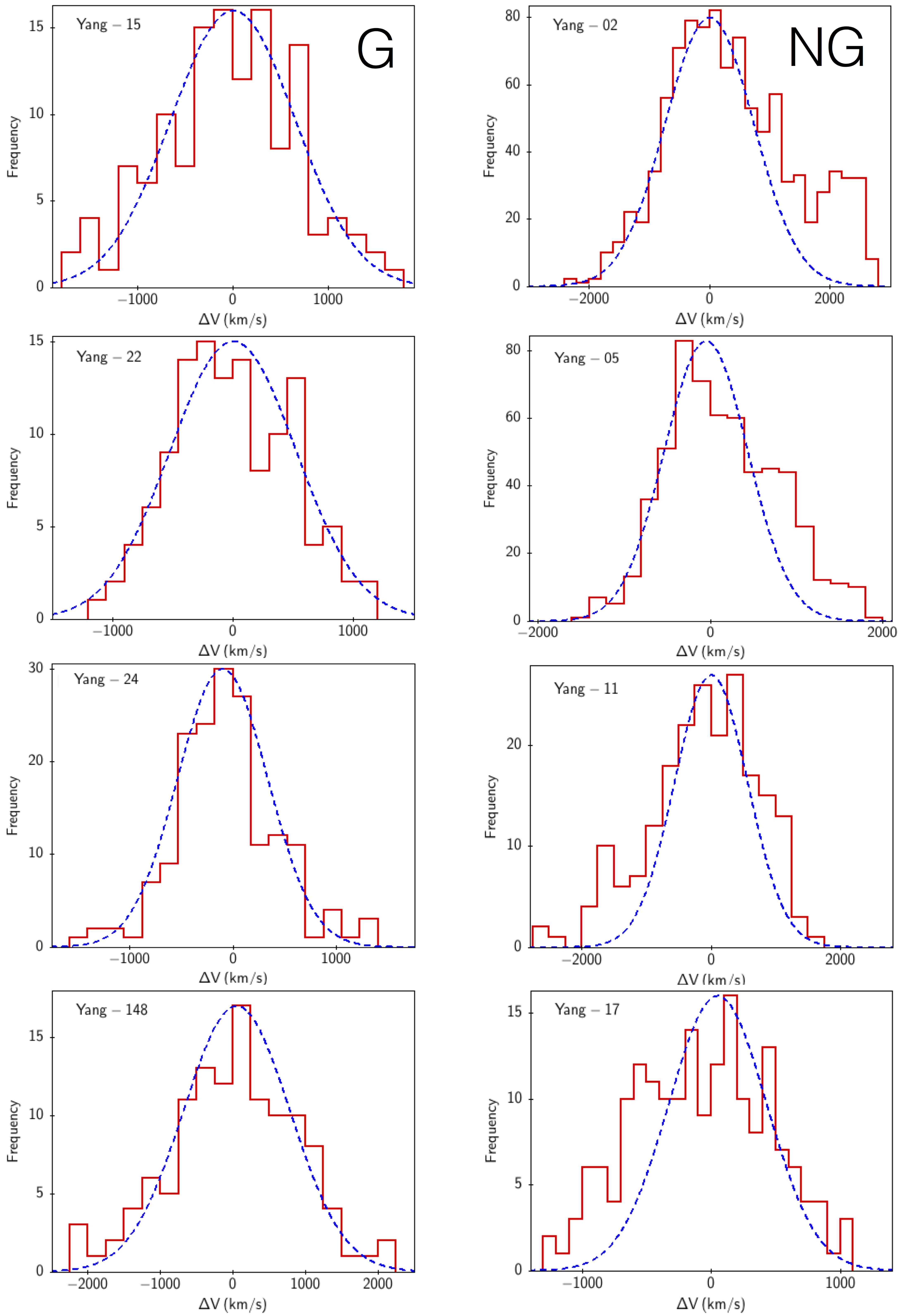}
		
		\caption{A few examples of velocity distributions of Yang groups studied here. The original ID of each Yang group is indicated in each panel in the upper left. The left column displays Gaussian systems, where the deviations from the gaussian, displayed in dashed blue line, are visually small. The right column exhibits Non-Gaussian groups and here we can clearly see significant deviations. We notice that this is only illustrative since in limiting cases we need a more robust and trustable method to separate G from NG and that is done by MCLUST and HD.\label{fig5}}

	\end{center}
	
\end{figure}

\begin{figure}[!ht]
	\begin{center}
		\includegraphics[scale=0.5]{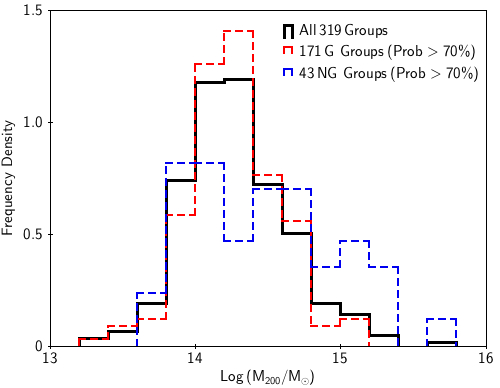}
		
		\caption{Comparison of the mass distributions according to the different dynamical stages of the groups. The median M$_{200}$ for NG groups is larger than for G ones by 0.22 dex. Similar tend is observed by \cite{RobertsParker2017}.\label{fig6}}

	\end{center}
	
\end{figure}

We find that 241 groups have gaussian velocity distributions (G) (241/319 $\sim$76\%), which is in agreement with the 70\% obtained by \cite{Ribeiro2013}, examining groups of the Berlind's sample. This is very reassuring since the method presented in \cite{Ribeiro2013} is similar but not quite the one employed in this work and Yang and Berlind samples are totally independent, even determined with distinct methods. Figure \ref{fig5} displays a few examples of velocity distributions of G and NG systems in the Yang sample, showing how well our gaussianity classification works. {This figure is only to illustrate the process. Subtle non-gaussianities have to be measured by the specific methods used here in this work and not established by visual impression}. In order to keep our analysis of the stellar populations of the galaxies in G and NG systems as meaningful (and consistent) as possible, we restricted our sample to groups for which the probability of the gaussianity, measured using bootstrap in the same way described in Section \ref{secDisMclustHD}, is higher than 70\%. Applying this criterion we end up with 171 G and 43 NG groups.  We measured how this limiting probability of gaussianity impacts on the total sample by comparing the mass (M$_{200}$) distribution of these two subgroups with the distribution for the whole sample of 319 systems (Figure \ref{fig6}). The permutation test \footnote[4]{Using the function permTS in R package under the library perm \citep[][]{fay2009}} is used to test the null hypothesis that two samples have identical probability distributions. We find that G systems have M$_{200}$ distributions similar to the total one (p-value = 0.19) while NG systems have M$_{200}$ distributions significantly different from to the total sample (p-value = 0.012). The observed discrepancy of the M$_{200}$ distributions of NG groups is more likely related to the asymmetry of the velocity distribution along the LOS, which may lead to an overestimation of the group's velocity dispersion and consequently its mass. This tendency of NG systems being more massive was already observed \citep[][]{Ribeiro2013, RobertsParker2017}. {Previously, \cite{Einasto12a} have found that richer and more luminous (and consequently more massive) clusters have larger amount of substructure, which is consistent to what we find in this work.} We note that this effect does not hinder our analysis, actually it points to a more fundamental problem of measuring virial mass using velocity dispersion, namely this scheme is only valid when the systems have a gaussian velocity distribution, which must be measured a priori.  

Another concern is related to the cutoff in richness when defining the groups from Yang sample. We impose a minimum number of twenty galaxies in a system (membership defined by Yang), to be included in the shift-gapper analysis and this translates into a cutoff in mass. From the M$_{200}$ and N$_{\rm R200}$ relation, where N$_{\rm R200}$ is the number of galaxies within R$_{200}$ with $\rm M_{\rm r} \leq -20.55$, we find that a mass cutoff of 10$^{14.0} \rm\hspace{1mm} M_{\odot}$ corresponds to N$_{\rm R200}$ = 20. This limiting mass reduces the sample size significantly, 143 G and 34 NG systems are left in the sample. 

Due to the close correlation of X-ray emission and mass for clusters of galaxies \cite[e.g.][]{ReiprichBohringer2002}, it is instructive to check, from X-ray cluster surveys, how much of this last sample has X-ray properties, in this case X-ray luminosity, L$_{\rm X}$, that might be useful as mass proxy. The two most recent X-ray cluster surveys with significant coverage are NORAS, \cite{Bohringeretal2000}, and REFLEX, \cite{Bohringeretal2004}, totaling a sample of 825 cluster with X-ray and spectroscopic data. Examining NORAS and REFLEX we look for the nearest (in projection) Yang groups in a search radius of 100 arc min and convert the angular separation between matched clusters to physical distances using the redshifts. Following \cite{Lopes2009} and \cite{Galetal2009}, we use as maximum physical distance the value of 1.5 Mpc and obtain only 22\% of our total sample match. However, when we look for Yang groups to the more heterogeneous BAX database (which is an online research database containing information on all galaxy clusters with X-ray observations to date), assuming the same criteria adopted previously, our match rate increases significantly to 58\% for the NG sample and 43\% for the G sample. Although our match rate has increased considerably, there is the possibility that these values have been affected due to selection of X-ray cluster samples being significantly biased low, $\sim$ 29\%, in favor of the peaked, Cool-Core objects \cite{Eckertetal2011}. 

\subsection{Measuring Skewness and Kurtosis - Searching for infall populations}

Visual inspection of the velocity distribution along the LOS of NG systems (Figure \ref{fig5}) shows clearly significant amount of skewness. In this Section, we quantify the deviation of the system's global velocity distribution along the LOS from a Gaussian using skewness and kurtosis. Skewness is related to the third, m$_{3}$, and the second m$_{2}$ (the variance) moments of the distribution \footnote[5]{m$_{2} = \frac{1}{n}\sum_{i=1}^n(x-\bar{x})^{2}, m_{3} = \frac{1}{n}\sum_{i=1}^n(x-\bar{x})^{3}$} and measures the asymmetric nature of the distribution -- negative or positive skewness indicates long left or right tail in the distribution, respectively.  Since we are always dealing with a sample instead of the whole population, the skewness can then be expressed following:

\begin{equation}
Skewness = \frac{\sqrt{n(n-1)}}{n-2} \frac{m_{3}}{m_{2}^{3/2}}
\end{equation}

\noindent where n is the number of data points \citep[see][]{cramer1997fundamental}. A more statistically meaningful measurement is the number of standard errors separating the sample skewness from zero and this is done dividing the Skewness by the standard error of skewness (SES) following the equation \citep[see][]{cramer1997fundamental}:

\begin{equation}
Z_{Skewness} = \frac{Skewness}{SES}
\end{equation}

\noindent where

\begin{equation}
SES = \sqrt{\frac{6n(n-1)}{(n-2)(n+1)(n+3)}}
\end{equation}

In the case where a distribution is symmetric, we can still measure the height and sharpness of the peak relative to the entire distribution, a quantity named kurtosis, defined by the fourth and second moments of the distribution \footnote[6]{m$_{4} = \frac{1}{n}\sum_{i=1}^n(x-\bar{x})^{4}$}. We express the sample kurtosis following also \citep[see][]{cramer1997fundamental} as

\begin{equation}
Kurtosis = \frac{n-1}{(n-2)(n-3)}\hspace{1mm}\Bigg[(n+1)\bigg(\frac{m_{4}}{m_{2}^{2}}-3\bigg) + 6\Bigg]
\end{equation}

\noindent where the term (m$_{4}$/m$_{2}$-3) is called excess kurtosis. Following the same reasoning as for Skewness, we write how many standard errors the sample excess kurtosis is from zero:

\begin{equation}
Z_{Kurtosis} = \frac{Kurtosis}{SEK}
\end{equation}

\noindent where

\begin{equation}
SEK = 2(SES) \sqrt{\frac{n^{2}-1}{(n-3)(n+5)}}.
\end{equation}

\begin{figure}[!ht]
	\begin{center}
		\includegraphics[scale=0.5]{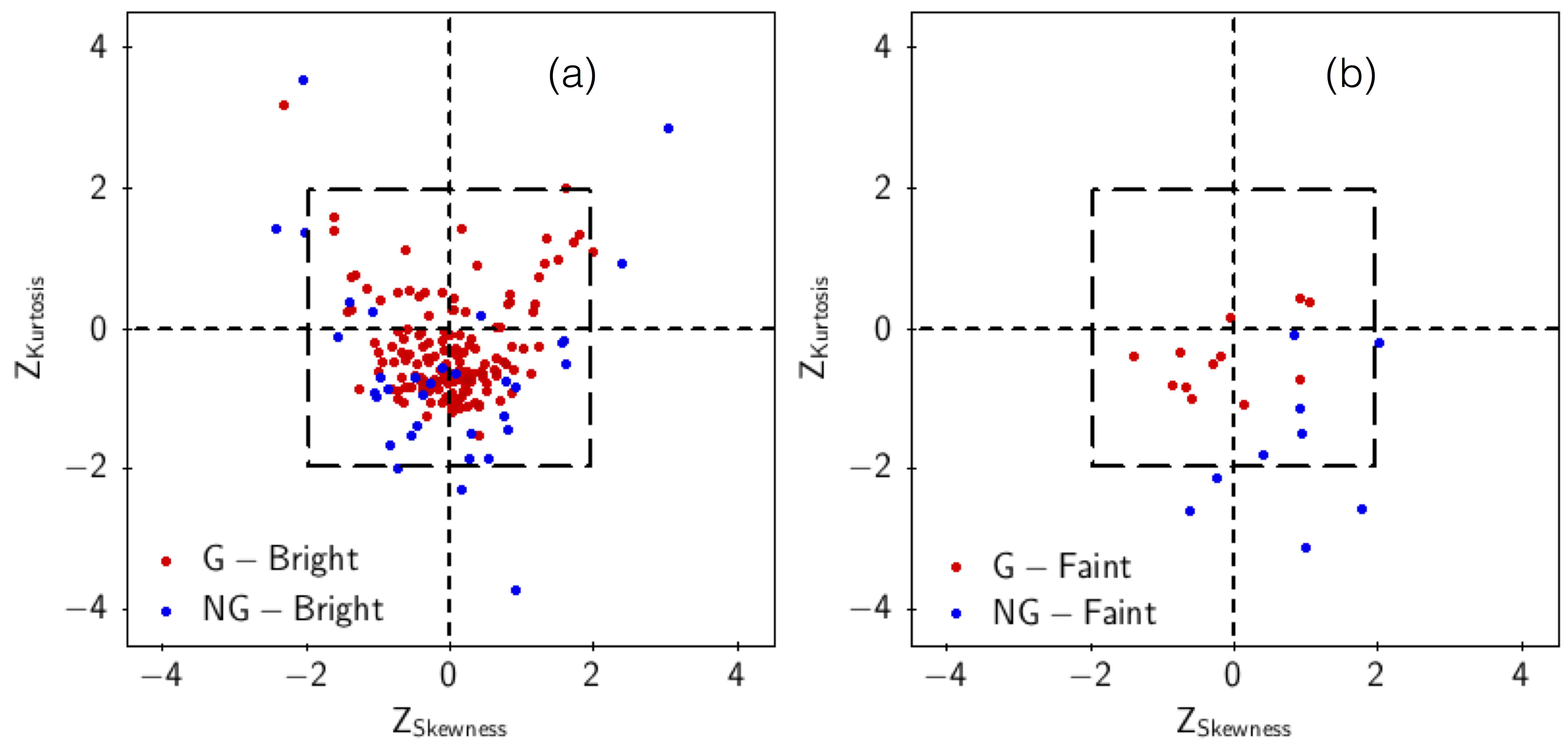}	
		
		\caption{(a) Excess of Skewness versus excess of Kurtosis for G and NG groups, using only bright galaxies. The box indicates the 95\% probability area. (b) the same as in (a) but using only faint galaxies.\label{fig7}}

	\end{center}
	
\end{figure}

Figure \ref{fig7} shows the measured skewness and kurtosis of the LOS velocity distribution of G and NG groups in the two magnitude regimes, bright (panel a) and faint (panel b). The dashed box indicates the region of a two-tailed test of Skewness and excess Kurtosis $\neq$ 0 at the 0.05 significance level ($\pm$ 1.96 for the Z$_{score}$ values). The test statistic indicates whether the whole population is probably skewed or platykurtic (or leptokurtic) \footnote[7]{platykurtic - excess kurtosis $<$0, means that in comparison with a gaussian, the studied distribution has its central peak lower and broader, and leptokurtic - excess kurtosis $>$0, means that it is higher and sharper} but not by how much - the bigger Z$_{score}$, the higher the probability. The box indicated in both panels of Figure \ref{fig7} is for 95\% probability. In Figure \ref{fig7}a we note that most of the data falls within the box where we cannot reach a firm conclusion on the skewness or kurtosis of the LOS velocity distribution. However, there is a systematic difference in Z$_{Kurtosis}$  with NG groups being more platykurtic than the G groups and negligible difference in Z$_{Skewness}$. The mean difference in Z$_{Kurtosis}$ between G and NG groups is $\sim$0.5. Also, there are 8 out of 34 (24\%) NG groups outside the box in contrast with 2 out of 143 (0.01\%) G groups, indicating that the velocity distribution of NG groups is more distorted {\it wrt} a gaussian than that of G systems. In Figure \ref{fig7}b, we compare again G versus NG groups looking at the faint galaxy population. It is very clear that NG systems have more negative Z$_{kurtosis}$ (mean $\sim$-1.67) than the G ones (mean $\sim$-0.43) with $\sim$50\% of the groups outside the box (5 out of 9). The mean Z$_{Skewness}$ for NG groups is around 0.77 while for G's is -0.15.  These results confirm that NG groups have LOS velocity distributions significantly different from a gaussian one. 

\subsection{What do we learn from the Projected Phase Space (PPS) ?}

It is a well known fact that properties of galaxies are affected by the environment trough which they pass during their life. In a simplified view, when a galaxy enters a filament experiences some pre-processing due to the increase in local density \cite{Porteretal2008} and eventually when it reaches a massive cluster will have its star formation history significantly changed. {The PPS carry a wealth of information on the dynamical state of the cluster. Therefore, in this section we investigate the relation between the stellar population properties of galaxies inhabiting different regions of the PPS}.

\begin{figure}[!ht]
	\begin{center}
		\includegraphics[scale=0.5]{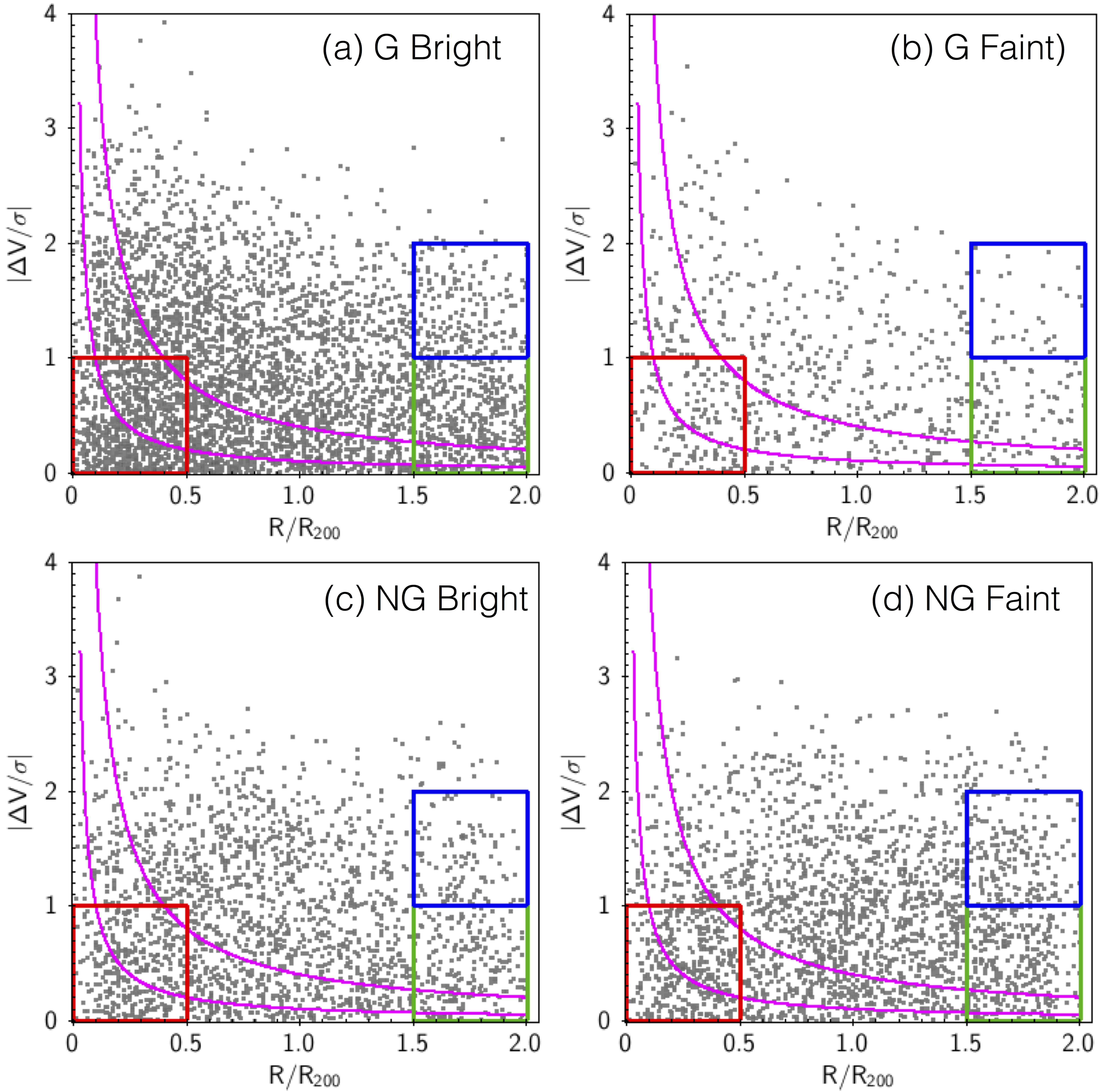}
		\caption{Stacked observed phase-space diagram for G and NG groups/clusters in our sample, separated by two different luminosity regimes.We followed  \cite{MMR11} to defined three specific regions of the PPS: Virial, Backsplash and Infall represented by the red, green and blue squares, respectively. The two magenta solid lines represent $\mid \Delta V/\sigma\mid \hspace{5pt} \propto  K/(R/R_{200}$), where K is equal to 0.1 and 0.4. \label{fig8}}	
	\end{center}	
\end{figure}

Figure \ref{fig8} displays the stacked projected phase-space diagram for G and NG systems separately considering the bright (panels a and c) and faint (panels b and d) regime of their luminosity functions. The peculiar velocity is normalized by the cluster velocity dispersion and the radial distance from the center of the system is normalized by the virial radius (R$_{200}$). We note that the number of galaxies in the faint regime, 3268, differs  significantly from that in the bright regime, 6506. From panel (b), we can clearly see that the difference is due to the faint component in G groups.  First, we have used an online Halo Mass Function calculator \citep[http://hmf.icrar.org/,][]{Murrayetal2013} to estimate the number of clusters in the $0.03 \leq \rm z \leq 0.04$ and masses $ > 10^{14.0} \rm\hspace{1mm} M_{\odot}$, regardless if the systems are G or NG. Different prescriptions for the Mass Function result in number of clusters between 22 \citep[][]{PS1974} and 41 \citep[][]{Bhatta2011}, which is consistent with the number of clusters we have in our sample, 31. This reinforces the fact that the difference we see between G and NG groups in the faint regime seems to be real. We count 761 galaxies in the faint regime of G groups compared to 2507 galaxies in NG groups. 

Considering that there is no obvious way of distinguishing galaxies in the PPS, we have used three different
approaches to define regions that may affect galaxy properties in distinct ways:

\subsubsection{Comparing PPSs using a kernel density estimation two-sample test}

We compare the PPSs defined for different environments in different luminosity regimes with the kernel density estimation  (KDE) test, which is a global non-parametric two-sample comparison test for 1 to 6 dimensional data, implemented in R under library $ks$ \citep[see][]{DGS12}. During the test, data are smoothed with a kernel function. The choice of this function is not crucial to the accuracy of kernel density estimators. The KDE test uses the general kernel of Wand \& Jones (1993) (see Appendix) and an optimized bandwidth matrix at each of the data points. After smoothing, two density distributions are achieved, $f_1$ and $f_2$,
which should be compared through the discrepancy measure,

\begin{equation}
T=\int \left[f_1(x) - f_2(x)\right]^2\;dx,
\end{equation}

\noindent here understood as the test statistic. \cite{DGS12} show that $T$ has a null distribution which is asymptotically normal, so no bootstrap resampling is required to compute an approximate p-value for the test.

We check the robustness of the KDE test results {\sl wrt} the variability of the data in the following way: in each comparison we run a bootstrap simulation creating 1000 random samples with replacement and each 
time we ask if the p-value is less than 0.05 (the significance level). Depending on the number of times the answer is yes or no we decide whether the samples are similar or not. For instance, in the comparison between the bright and faint samples of G systems, we find that in 0 out of 1000 cases the p-value is less than 0.05, indicating that these two samples are statistically similar. Notice from Table \ref{tablePPSkde} that, GF X NGF and NGB X NGF are all statistically different, while GB is statistically similar to NGB. These results reinforce, once again, that the discrimination between G and NG does not result from any methodological detail and seems to genuinely represent a physical difference, specially when we focus on the faint component.

\subsubsection{Ad Hoc Definition of Regions of the PPS}

The second test invokes arbitrary definitions of three specific regions of the PPS: Inner region (R/R$_{200} < 0.5$), Intermediate region ($0.5 < {\rm R/R_{200} }< 1.0$), and Outer region (R/R$_{200} > 1.0$). Also, we distinguish between Low velocity (LV) ($| \Delta V / \sigma | < 0.5$) and High velocity (HV) ($| \Delta V / \sigma | > 0.5$). Table \ref{tablePPSanalysis} summarizes the statistics for the regions. Median values are presented for Log M$_{stellar}$, Age, and [Z/H], as well as the fraction of galaxies in each region and the p-values when comparing LV and HV subspaces.  We can summarize our findings with this type of analysis of the PPS in the following way:

\begin{itemize}
	\item{The first point to highlight when examining the G-BRIGHT results is that LV and HV galaxies in the central regions  are statistically different as far as Log M$_{stellar}$, Age, and [Z/H] are concerned. LV galaxies are more massive, older and have higher metallicity than HV galaxies. In the intermediate region, the differences in Age and [Z/H] remain, but not in Log M$_{stellar}$, while in the outer regions we did not observe significant differences between LV and HV galaxies. The fraction of LV galaxies does not change from inner to outer regions and the fraction of HV galaxies shows a slight increase toward the center.}
	\item{Extending the analysis to G-FAINT, we find no significant differences between LV and HV galaxies (see p-values) in any clustercentric distance, although a small gradient in Age and [Z/H] occurs for LV and HV galaxies.}
	\item{Again, for NG-BRIGHT we do not find significant differences between LV and HV galaxies, with only is a small trend of older Age towards the center (mainly for LV galaxies).}
	\item{The NG-FAINT subsample is where we find more significant differences. In the central and intermediate regions LV galaxies are older than the HV ones. In the intermediate region we find that LV galaxies are significantly more metal rich than the HV ones. In the outer regions, Log M$_{stellar}$, Age, and [Z/H] are indistinguishable;}
\end{itemize}

\subsubsection{Defining Regions of the PPS Based on Cosmological Simulations}

As an independent check on how the properties of galaxies vary over the PPS, we defined, instead of specific regions as in the preceding subsection (which are arbitrary), different regions indicated by results obtained through the analysis of cosmological simulations \cite{MMR11}.  In Figure \ref{fig8} we show three main regions of interest in the PPS that may be reflecting the accretion epoch: a) the virial region (in red, hereafter denoted by VIR) is likely to be dominated by galaxies which participated of the cluster core formation at early times; b) the backsplash region (in  green, hereafter denoted by BS) where galaxies have passed through the cluster core once and are heading out of the cluster; and c) the infall region (in blue, hereafter denoted by INF) populated by galaxies that have been accreted to the cluster from the surroundings. It is important to note that these regions defined in  \cite{MMR11} are not arbitrary (see their Table 2), they were chosen in order to maximize the fraction of VIR, INF and BS particles of the stacked mock cluster \citep[from][]{BorganiEtal2004} in cells of projected phase space.
\cite{Omanetal2013} have shown that although we see a lot of structure in the radial phase-space (radial velocity versus radial position) that is lost when we exam the PPS (projected LOS velocity versus projected radial position), the latter allows better separation between VIR, BS and INF galaxies. These three locations are well separated in radial phase-space diagram \citep[e.g.][]{MMR11}. We examine the stellar population properties in these three regions aiming to find a relation between the star formation history and the environment, where here we interpret environment not only as G versus NG but also which region of the phase-space the galaxy is. 

\begin{figure}[!ht]
	\begin{center}
		
		\includegraphics[scale=0.5]{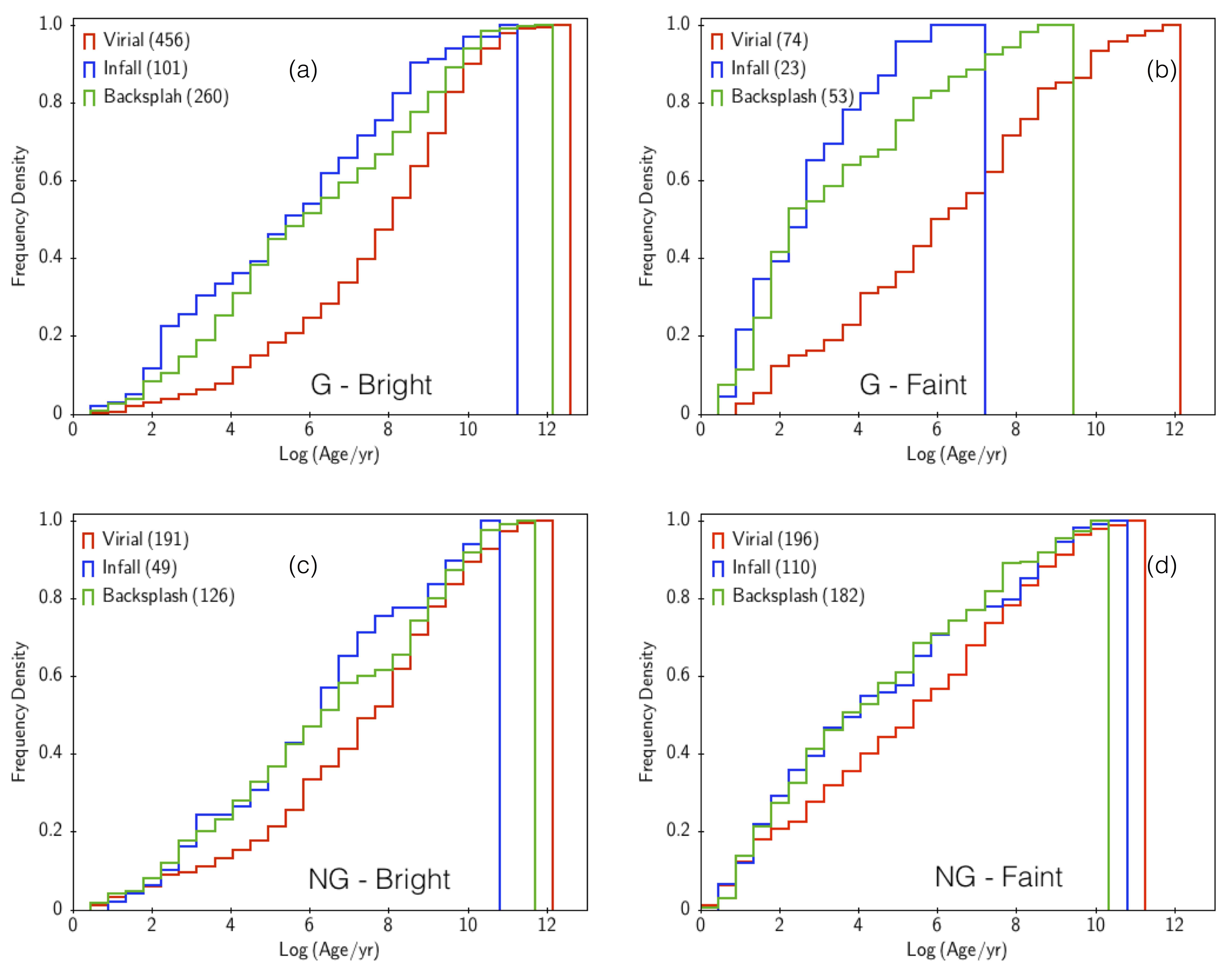}
		\caption{Cumulative distribution of age in different regions of the phase-space diagram, as described in Figure \ref{fig8}. The colors are the same as in Figure \ref{fig8}, red for Virial, green for Backsplash and blue for infall. Number of galaxies in each region and in each case (G versus NG and Bright versus Faint) are indicated in parenthesis on the upper left of each panel. We call the reader's attention to the remarkable difference between G-Faint and NG-Faint.\label{fig9}}	
		
	\end{center}	
\end{figure}

Figure \ref{fig9} displays the cumulative distribution of age in three distinct regions of the phase-space. We compare the distributions by using the permutation test. Table \ref{tablepvalues} presents the comparisons between VIR, INF and BS for a given environment, G or NG. As we did previously, we test the null hypothesis that two samples have identical probability distributions. In what follows we impose a significance level of 5\%, namely if the p-value is less than or equal to the chosen significance level (0.05), the observed data is inconsistent with the null hypothesis, meaning that the two distributions are statistically different. In panel (a), we see that the cumulative distribution of the age of the galaxies in the VIR region differs significantly from those in the BS and INF regions while we do not see any significant difference between the age distributions of galaxies in BS and INF. If we ask  which fraction of the galaxies in each region have ages less than 7 Gyrs (the median age of all bright galaxies in G systems) we find that in the VIR is $\sim38\%$, in the INF $\sim60\%$ and  in the BS $\sim70\%$. These numbers show unequivocally that in G systems, bright galaxies in the BS and INF regions are significantly younger than those in the VIR region. In panel (b), we extend the comparison taking into account only the faint galaxies and the result is somewhat different - age of galaxies in the VIR region is significantly different from those in the INF region but similar to those in the BS region, while the age distribution of galaxies in BS and INF are statistically similar. It is important to note that although we considered the age distributions of galaxies in VIRand BS similar the significance (0.068) is quite close to the limiting value we used (0.05). In this case we find that $\sim25\%$ of the galaxies in the VIR region have ages less that 4 Gyrs (the median age of all faint galaxies in G systems), while in the BS region this number is $\sim65\%$ and in the INF region is $\sim76\%$. There are no galaxies in the INF (BS) region older than 7 (10) Gyrs. We can clearly see that faint galaxies, with BS and INF orbits, in G systems are very different from the VIR ones, manifesting a significant environmental effect. Panels (c) and (d) are similar to the panels (a) and (b) but for the NG systems. The same qualitative results were found, namely when examining the bright galaxies we find that those in BS and INF regions have similar age distributions and both are statistically different from those in the VIR region. However, it is noticeable that the distributions are closer to each other than in the case of G systems. The fraction of bright galaxies with ages less than 7 Gyrs is $\sim43\%$ in VIR, $\sim56\%$ in BS and $\sim63\%$ in INF. These fractions are much closer to each other compared to the ones for bright galaxies in G systems. For the NG systems the difference {\it wrt} to G systems is even larger, the fraction of faint galaxies with ages less than 4 Gyrs is $\sim40\%$ in VIR, $\sim50\%$ in BS and $\sim55\%$ in INF. Comparison of panels (b) and (d) shows, even visually, how the star formation history of faint galaxies in NG systems seems to be very different from the faint ones in G systems. 

\begin{figure}[!ht]
	\begin{center}
		
		\includegraphics[scale=0.5]{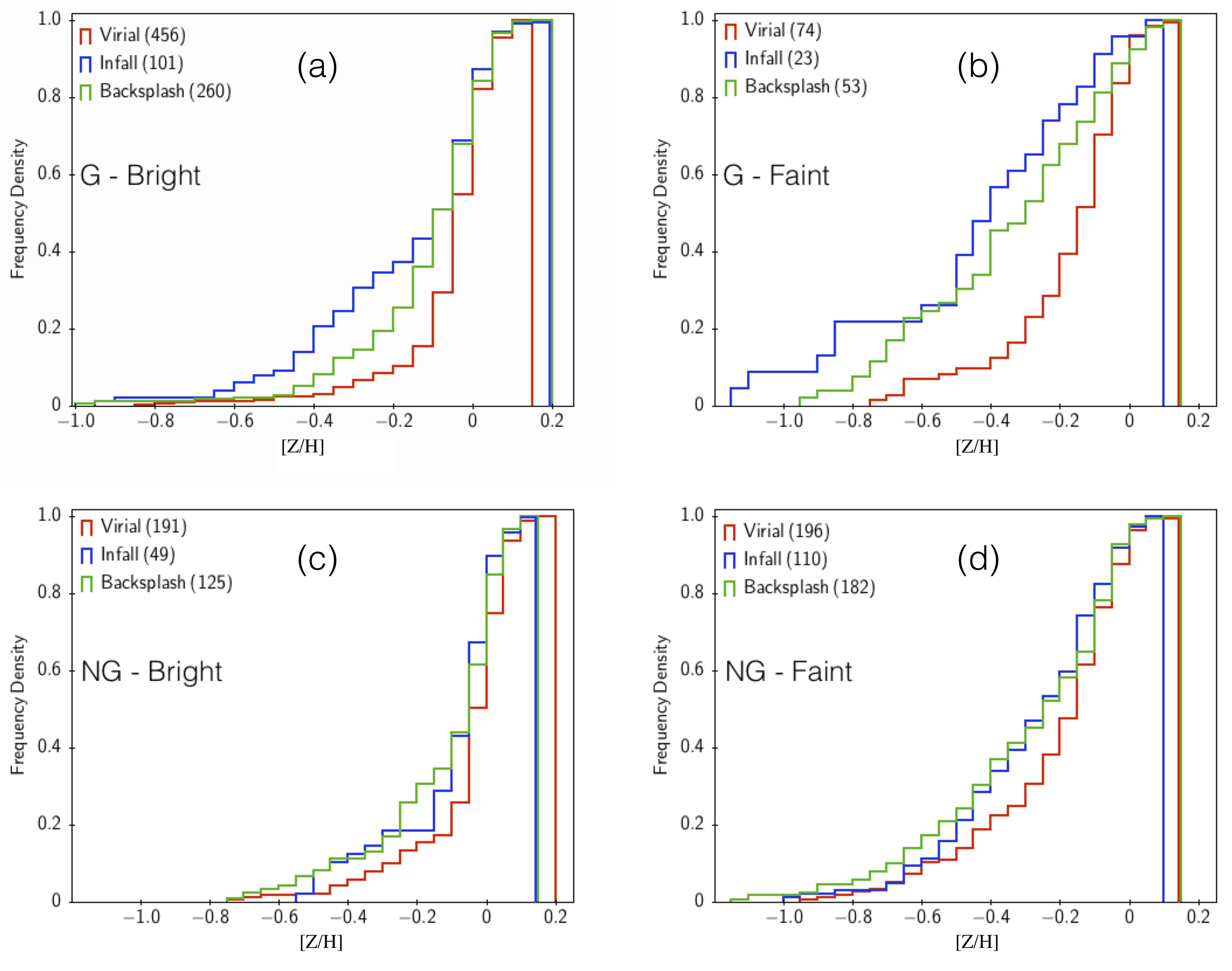}
		
		\caption{The same as in Figure \ref{fig9} but for metallicity.}\label{fig10}	
		
	\end{center}	
\end{figure}

Figure \ref{fig10} exhibits the cumulative distribution of metallicity in the same three distinct regions of the phase-space as presented  in Figure \ref{fig9}  for the age distribution. Comparison of the distributions in the VIR, INF and BS regions, based on the permutation test, is also presented in Table \ref{tablepvalues}. Keeping the same significance level of 5\%, we find that for bright galaxies in G systems all three regions exhibit significantly different [Z/H] distributions. As for the faint galaxies in G systems, the [Z/H] distribution in the VIR region is significantly different from INF and BS, while these two regions present similar [ZH] distributions. Regarding the bright galaxies in NG systems the situation is different. In this case, the [Z/H] distributions of galaxies in VIR and INF are similar, as well as those in the INF and BS. However, galaxies in the VIR and BS regions have [Z/H] distributions significantly different. The faint galaxies in NG systems have the same behavior as bright galaxies as far as [Z/H] distributions are concerned, which can be seen from Table \ref{tablepvalues}. Another comparison worth doing is between bright and faint galaxies in each environment, G and NG, and in each region, VIR, BS and INF. All comparisons have displayed a p-value of 0.002, indicating that bright and faint galaxies have age and [Z/H] significantly different regardless they are in G or NG and regardless the type of orbit they are in. This is an important result that will be further explored in Section \ref{Secdiscussion}.

\begin{figure}[!ht]
	\begin{center}
		
		\includegraphics[scale=0.5]{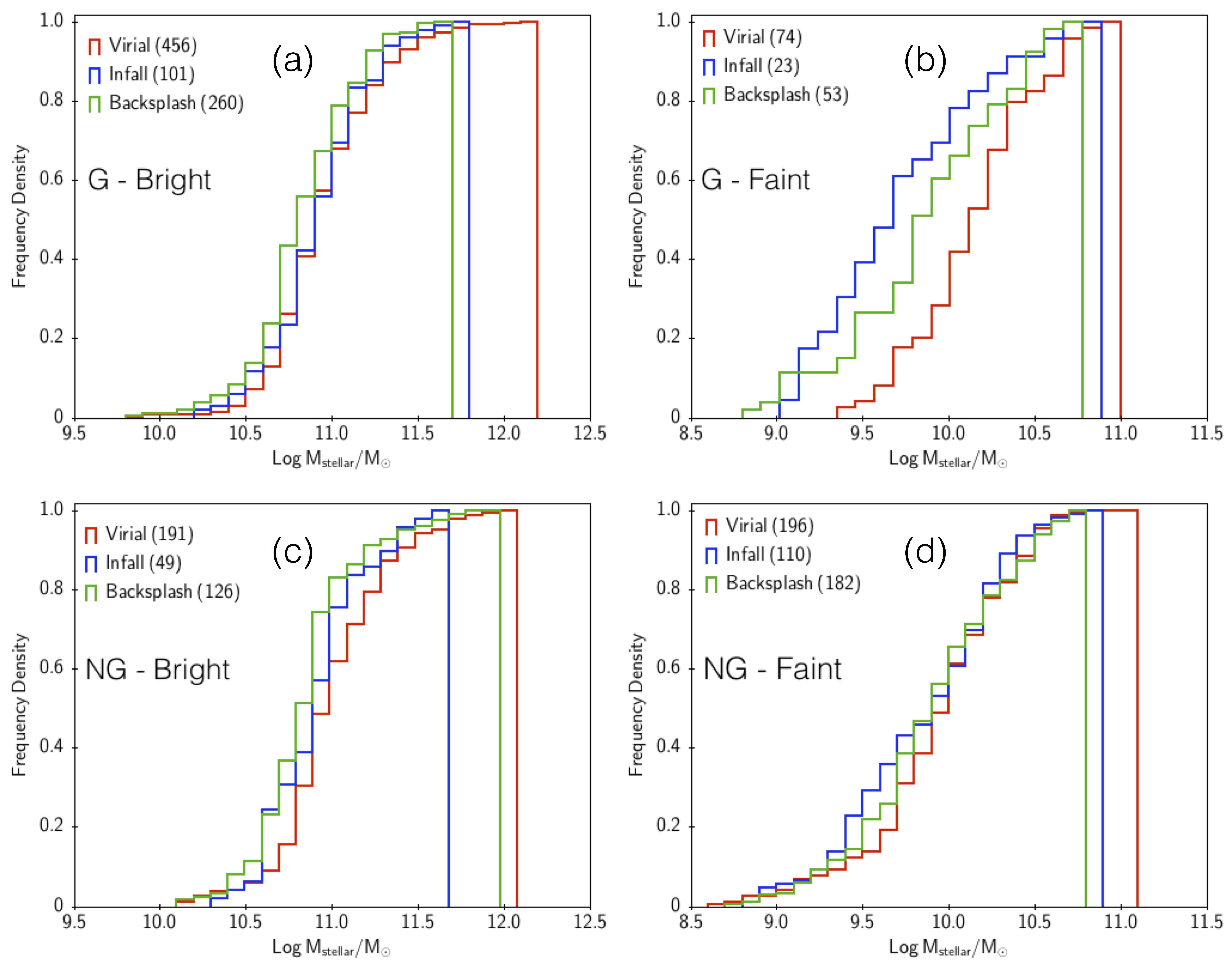}
		
		\caption{The same as in Figure \ref{fig9} but for stellar mass.}\label{fig11}	
		
	\end{center}	
\end{figure}

In Figure \ref{fig11} we present the cumulative distribution of stellar mass in the three distinct regions of the phase-space as in Figures \ref{fig9} and \ref{fig10} (See Table \ref{tablepvalues} for the permutation test results). Looking at the bright galaxies in G systems (panel a), we find a significant difference between the distribution of stellar masses of galaxies in the INF and BS regions in comparison with that of galaxies in the VIR region, in the same way as we found for Age. The high-end stellar mass of bright galaxies in the VIR region of G systems is roughly 0.5 dex higher than those in the INF and BS regions. In panel (b) faint galaxies in G systems are compared as far as the stellar mass distribution is concerned and here only VIR and INF are different, the others, VIR versus BS and INF versus BS are statistically similar. However, we should note that comparison between VIR and BS is only slightly above the limit of 0.05 used here. Panels (c) and (d) do similar comparisons as presented in panels (a) and (b) except that in this case we consider NG systems. As attested by the results presented in Table \ref{tablepvalues}, all distributions are statistically similar, using the same significance level of 5\%.

\subsection{How do the Stellar Population of galaxies respond to the Environment ?}

\begin{figure}[!ht]
	\begin{center}
		
		\includegraphics[scale=0.5]{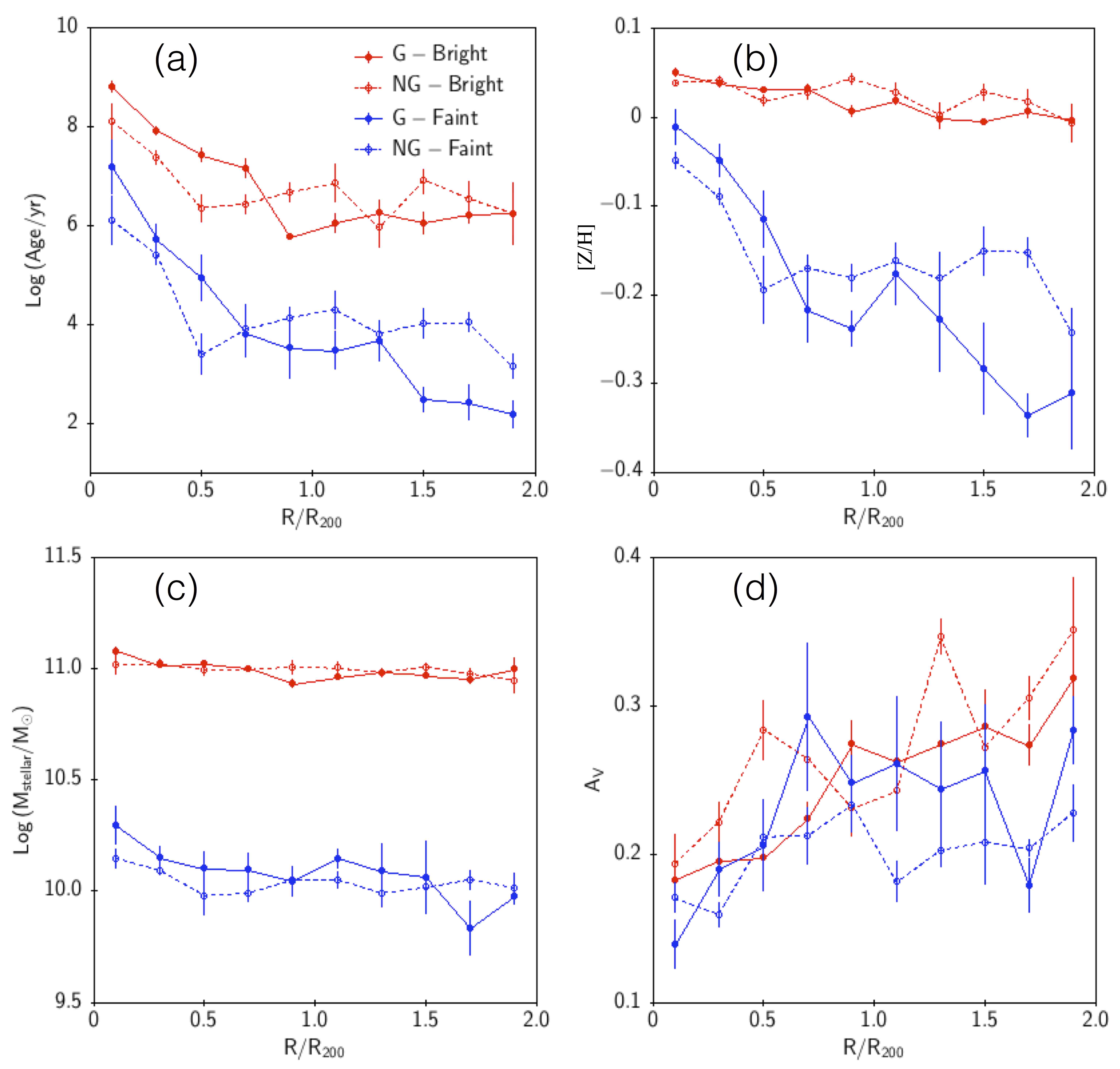}
		\caption{Stellar population parameters, Age, [Z/H], M$_{stellar}$ and A$_{V}$ as a function of the cluster-centric distance normalized by R$_{200}$. Red and blue solid lines indicate Bright and Faint magnitude regimes, respectively. Environment is represented by solid (G) and dashed (NG) lines.  }\label{fig12}	
		
	\end{center}	
\end{figure}

In this Section we explore a different way of probing the environmental effect, namely by measuring galaxy properties as a function of the distance from the center of the cluster. As we can clearly see from Figure \ref{fig8}, for the range $1.5\leq {R/R_{200}} \leq 2.0$ we have a mixture of galaxies with infall and backsplash orbits and they seem to have significantly different metallicity distributions, for instance. Therefore, to further study how the stellar population of galaxies in groups depend on the environment, we investigate four quantities of interest as a function of the distance from the center of the cluster, normalized by R$_{200}$: Age of the stellar population weighted by luminosity expressed in Gyr. This parameter reflects more specifically the last star formation episode in the galaxy rather than a global age; metallicity, [Z/H], in solar units; stellar mass, $\rm M_{stellar}$ in M$_{\odot}$; and internal extinction, A$_{\rm V}$. Figure \ref{fig12} displays all these quantities. The profiles were established in bins of R/R$_{200}$ = 0.2 and in each bin we measure the median and Q-sigma, a robust estimator of the standard deviation (Q-sigma = 0.7415*(Q75-Q25), where Q25 and Q75 are the quartiles of the distribution). In panel (a) we see that a certain trend is present for bright as well as for faint galaxies regardless of the G or NG characterization of the velocity distribution and that is for $R/R_{200} \leq 0.75$ bright galaxies in G systems are older than bright galaxies in NG ones by 0.71 Gyr. For $R/R_{200} > 0.75$ we see the opposite  trend by 0.56 Gyr. Examining the faint galaxies the behavior is the same with differences in age of 0.70 and 0.90 Gyr, respectively. In panel (b), we see that metallicity behaves somewhat similarly. Within $R/R_{200} \leq 0.75$, bright galaxies in G systems are slightly more metal rich than their counterparts in  NG systems, $[Z/H]_{G} - [Z/H]_{NG}$ = 0.01 and for $R/R_{200} > 0.75$, $[Z/H]_{G} - [Z/H]_{NG}$ = -0.02. As for the faint galaxies we notice that for $R/R_{200} \leq 0.75$ the difference $[Z/H]_{G} - [Z/H]_{NG}$ = 0.04, namely in the central region faint galaxies in NG systems are significantly more metal poor than faint galaxies in G systems, while in the $R/R_{200} > 0.75$ region $[Z/H]_{G} - [Z/H]_{NG}$ = -0.06. In other words, in the outskirts there is a large difference in metallicity  when we compare faint galaxies in G and NG systems, evidencing a significant difference in stellar population properties between G and NG, as far as faint galaxies are concerned. The same effect is present in Age but not as significant as for faint galaxies. Panel (c) compares the stellar mass of bright and faint galaxies in G and NG and we see that for bright galaxies there are no differences between G and NG groups - for $R/R_{200} \leq 0.75$ we have $\Delta M_{stellar}$ = 0.05 dex and for $R/R_{200} > 0.75$, $\Delta M_{stellar}$ = -0.07 dex. When we look at the faint population, once again the situation is significantly different. For $R/R_{200} \leq 0.75$ we have $\Delta M_{stellar}$ = 0.28 dex and for $R/R_{200} > 0.75$, $\Delta M_{stellar}$ = -0.05 dex. Here we see faint galaxies having significantly different $M_{stellar}$ only in the central regions, in NG systems they are less massive than in the G ones. In panel (d) we exhibit internal extinction, as a function of the clustercentric distance. As we can clearly see the variation (measured by the standard deviation in each bin) is very large, preventing any reliable comparison between bright and faint galaxies in G and NG systems. The only global trend we can see is that A$_{\rm V}$ increases as we probe the outskirts on a cluster, which is expected as a consequence of the morphology density relation.

\section{Discussion\label{Secdiscussion}}

Environment plays a major role in determining how galaxies evolve. Since \cite{DRESS80}, we learned that galaxies in high galactic density are different {\it wrt} those in the low density regime. In this paper, we investigate the galaxy properties in clusters, from the center to the outskirts spanning roughly seven orders of magnitude in luminosity surface density. In the last fifteen years several contributions lead to the indication that clusters can be modeled simply as a virialized component dominated by old galaxies plus a quasi-equilibrium one mainly constituted by younger galaxies \citep[e.g.][]{Carlberg1997,ELLIN01}. The later, results more likely from recent accretions from filaments which may alter the galaxy properties significantly before they mix with the older and virialized population. 

In this study, we investigate the relationship between stellar population properties and cluster environment. To define environment, we considered two independent ways of measuring the gaussianity of the velocity distribution (MCLUST and HD) and attributed a probability to it. We have used simulated data to assess the limits of applicability of the methods employed. This is quite an improvement {\it wrt} the methodologies based on more traditional normality tests  \citep[see][]{Ribeiro2013}. We then study the groups in the Yang's catalog and essentially HD and MCLUST agree reasonably well, 75\% when $\delta \ge$1.7, reinforcing their strength in distinguishing G from NG very accurately as long as the probability of being G or NG is high (larger than 70\%, for instance). In Figure \ref{fig5} we can clearly see how the G groups are more symmetric than the NG ones, which present significant tails in the distributions. 

Although the deviations in the velocity distribution are clearly seen, a more quantitative measure is needed. Here, we have measured the excess of skewness (Z$_{Skewness}$) and kurtosis (Z$_{Kurtosis}$). Figure 7a shows a significant difference between G and NG when taking into account only bright galaxies, indicating that the separation between G and NG is not fortuitous.  NG groups have a very negative Z$_{Kurtosis}$ in comparison with G groups. But the most striking result is when we examine the faint galaxies in both environments. Here we estimate an average Skewness and Kurtosis and compare directly to the results obtained by \cite{VijayaraghavanEtal2015}. They run simulations to study how are dwarf galaxies affected when a group infall to a cluster. Their findings are very elucidating when compared to ours. First, in their case, the velocity distribution of dwarf galaxies have a high positive Skewness ($\sim1.0$) in the first pericentric passage and a low negative Skewness in the second passage ($\sim$-0.3). The variation of Skewness as a function of time does not seem to depend on the mass of the group and cluster and also on the light of sight we measure the velocity distribution. For comparison, we measure a median Skewness of 0.17$\pm$0.16 for the faint galaxies in NG groups (here we measure Skewness and not Z$_{Skewness}$ to be compatible with their results), which is consistent with the picture where these dwarf galaxies are seen right before or after the first pericentric passage. As far as Kurtosis is concerned, \cite{VijayaraghavanEtal2015} show that the variation with time is strongly dependent on the mass of the group and the cluster and overall there is a peak with positive Kurtosis ($\sim 1.2$) during the first pericentric passage and then a monotonic increase with time. It is interesting to note that before the first pericentric passage, Kurtosis has reached  its minimum value ($\sim -0.5$). In comparison, we measure a Kurtosis of -0.66 $\pm$ 0.57. Based on both measures, Skewness and Kurtosis, we conclude that faint galaxies in NG groups are mainly infalling for the first time in the cluster. Obviously, this result should be seen in average for the family of NGs but it is noticeable that 6 out 9 NG systems have Kurtosis $<$ -0.5, strongly supporting the view that faint galaxies in these systems are in the very early stage of infalling, before the first pericentric passage \citep[see Figure 8a of][]{VijayaraghavanEtal2015}.

Comparison of the PPS using the whole 2D distribution indicates that faint galaxies of G and NG systems are distributed very differently (see Table \ref{tablePPSkde}). There are far more faint galaxies in NG than in G systems. This is further supported by the fact that in NG groups bright and faint galaxies are also distributed differently, which is not the case for G groups. This last finding is very reassuring that G groups have bright and faint galaxies distributed similarly in the PPS because they may have reached dynamical equilibrium. These trends may be associated to a higher infall rate in NG groups and if this is the case we should find signs of pre-processing. With that in mind, we examined the cumulative distribution of Age, [Z/H], and $\rm M_{stellar}$ and found that for G systems there are no faint galaxies in the INF (BS) region older than 7 (10) Gyrs, possibly manifesting the morphology density relation. As for the NG systems, on the contrary, we find that the age distribution for all three distinct orbit classes are statistically similar, which may be interpreted as a higher infall rate of galaxies into the NG groups. In this sense, NG systems are the ones with more disturbed velocity distribution and the stellar population properties are well mixed. This reinforces how the dynamical state is intimately related to the average stellar population. When we examine the metallicity distribution we find essentially the same qualitative result but one striking feature is noted - there is an obvious excess of more metal rich galaxies in the faint systems of NG groups than their G counterparts. Also, there is an excess of higher stellar mass galaxies in the NG-Faint than in the G-Faint groups. Both results may be related to preprocessing mechanism and agrees well with results from \cite{RobertsParker2017}. An important feature that shows the possible action the pre-processing mechanism is the way Age and [Z/H] vary with clustercentric distance. For R$\leq$0.75R$_{200}$ bright galaxies in G groups are older than the ones in NG groups, while for R$\geq$0.75R$_{200}$ is the opposite, bright galaxies in NG groups are older than the ones in G groups. The same trend is observed for the faint galaxies. Regarding metallicity we see almost the same behavior, although in the central regions (R$\geq$0.75R$_{200}$) bright galaxies in G groups are only slightly older the ones in NG groups. In summary, these profiles show that in the outskirts of NG groups, galaxies are older and more metal rich than galaxies in the outskirts of G groups. Notice also, that stellar masses have very similar distributions in G and NG systems, indicating that the way gas is converted into stars has an efficiency independent of the environment, which reproduces quite well the result obtained by\cite{CarolloEtal2013}. All these findings based on the PPS are directly related to the way quenching is affecting the galaxy stellar population properties. As pointed out by \cite{PaccagnellaEtal2016}, environmental mechanisms should affect star forming galaxies as they move from the field to groups/clusters, depending critically on the quenching timescale. This is in agreement with the fact that we find Ages around 4 Gyrs for NG systems even at 1.5R$_{200}$. Considering that infall may happen at much larger radii ($\sim$4-5$_{200}$) quenching seems to be slowly progressing even at 1.5R$_{200}$ without turning the stellar populations into a complete passive mode. In the LoCuSS project (Local Cluster Substructure Survey), two specific papers \citep[][]{HainesEtal2013, HainesEtal2015} find results consistent with galaxies being slowly quenched upon arrival in the cluster, their SFRs declining exponentially on timescales in the range 0.7$-$2.0 Gyr. Also, \cite{SmithTaylor2008} present a study connecting substructures in cluster cores with their assembly histories, showing  that clusters with higher fraction of substructures (the ones probably having more disturbed velocity distributions) are formed more recently.  These results are complementary to ours, especially when we show that NG systems may be experiencing a higher infall rate, and the fact that in  the inner regions of G groups galaxies are older and more metal rich than in the inner regions of NG ones.

We also compare LV and HV galaxies between the G and NG environments. An important outcome of this analysis is to verify that HV galaxies are comparable in both environments, while LV galaxies are older in the G-bright sample (up to R$_{200}$) than in the NG-Bright sample; and LV objects are younger and exhibit lower metallicities in the G-faint sample (at R$>$ R$_{200}$) than in the NG-faint sample. Taken together, these results suggest environmental mechanisms acting on galaxies, especially if we understand that LV objects are those which have been in the cluster environment for the longest time. This is in agreement with the fact that significant differences always occur indicating more evolution in LV objects, strengthening the idea of environmental effects acting on these galaxies. On the other hand, the presence of older LV objects with higher metallicities in the NG-faint sample (for R$>$R$_{200}$) than in the G-Faint sample possibly reflects some pre-processing effect which would be occurring only in the surroundings of NG systems, again in agreement with  \cite{RobertsParker2017}. 

\section{Summary\label{Secsummary}}

In this contribution, we present a new way to characterize the velocity distribution of galaxies in a group/cluster as Gaussian or Non-Gaussian. We discuss the limits of applicability of two independent methods to establish gaussianity and their reliability. We then study massive groups in the Yang sample to investigate the relation between galaxy properties and the non-gaussianity of the velocity distribution. Below we summarize some of the main results of this paper.

(i) We investigate two independent methods of measuring the degree of gaussianity of a velocity distribution. MCLUST identifies multiple gaussians in the distribution while Hellinger Distance measures how far from a gaussian the distribution is. We show that although both methods work similarly when applied to the Yang's catalog of groups, HD outperforms MCLUST specially in the range of smaller $\delta$.

(ii) Examining 319 groups of the Yang's catalog, we find that 76\% have gaussian velocity distributions, in agreement with \cite{Ribeiro2013} who found 70\%.  NG groups are more massive than the G ones by 0.22 dex, which confirms previous finding \citep[][]{Ribeiro2013, RobertsParker2017}. This is an important issue which may have a consequence in the mass estimate based on velocity dispersion, specially for high-z clusters where we expect to have more NG than G systems \citep{Ribeiro2013}. Also, by inspecting the BAX database we find that 58\% (43\%) of the NG (G) groups have X-ray detection.

(iii) We measured the deviation of the systemÕs global velocity distribution along the LOS from a Gaussian, by estimating Skewness and Kurtosis. In general we find that the velocity distribution of NG groups is more distorted {\it wrt} a gaussian than that of G systems. Our results in comparison with simulations done by \cite{VijayaraghavanEtal2015} indicates that faint galaxies in the outskirts of NG groups are infalling for the first time previous to the first pericentric passage.

(iv) Examining the PPS, we find that G groups have bright and faint galaxies distributed similarly, indicating dynamical equilibrium and the fact that in the case of NG groups bright and faint galaxies have different PPS distributions reinforces the fact that NG systems may be experiencing a higher infall rate. No surprising that NG groups are the ones with more disturbed velocity distributions.

(v) There is a clear excess of more metal rich galaxies in the faint galaxies belonging to NG groups compared to  the G ones. Also, faint galaxies in the NG groups have higher stellar masses than the faint galaxies in the G groups. These trends are well explained by preprocessing, fully consistent to what is found by \cite{RobertsParker2017}.

(vi) Analysis of the stellar population content as a function of the distance from the center of the group lead us to conclude that regardless whether galaxies are bright or faint, in the inner regions of G groups they older and more metal rich than in the inner regions of NG ones. Also, in the outer regions of NG groups, galaxies are older and more metal rich than galaxies in the outskirts of G groups, regardless of their luminosity. This is all further evidence of preprocessing acting in the surroundings of NG groups.

\section*{Acknowledgments}

\noindent RRdc wishes to thank Dr Roy Gal, Gary Mamon and Francesco La Barbera for stimulating discussions on this topic over the years. We would like to thank the anonymous referee for the very helpful report which helped to improve the manuscript significantly. RRdC and RRR acknowledge financial support from FAPESP through a grant $\#$2014/11156-4. DHS acknowledges the financial support from CNPQ scholarship $\#$140913/2013-0 and $\#$201636/2015-8. ALBR  acknowledges the financial support from CNPq grant 309255/2013-9.
\bibliography{references}
\section*{Appendix}

\vspace {0.5cm}

\section*{Non-parametric test to compare two-dimensional distributions\label{apendixTest}}

The most used non-parametric two-sample tests for one-dimensional data are the  Kolmogorov-Smirnov and Anderson-Darling tests. These tests, however, cannot be applied in two or higher dimensions, because there is no unique way to order the points so that distances between two distribution functions can be computed \citep[see][]{feigelson2012modern}. Alternatively, kernel smoothing is a widely used computational technique for density estimation due to its intuitive construction and interpretation \citep[][]{simonoff2012smoothing}. Thus, it is an ideal basis for non-parametric  density-based testing. Kernel-based tests have been developed with other discrepancy measures \citep[][]{martinez2008k}, but all rely on computationally intensive resampling methods to compute the critical quantiles of the null distribution. A more efficient method with respect to computational complexity is the so-called ``black-box'' comparisons of multivariate data (Duong, Gould \& Schauer 2012). The algorithm transforms data points into kernels and develop a multivariate two-sample test that is nonparametric and asymptotically normal to directly and quantitatively compare different distributions. The asymptotic normality bypasses the computationally intensive calculations used by the usual resampling techniques to compute the p-value. Because all parameters required for the statistical test are estimated directly from the data, it does not require any subjective decisions. We give now a brief description of the method.

Let X$_{1}$,X$_{2}$,...,X$_{n_{1}}$ and Y$_{1}$,Y$_{2}$,...,Y$_{n_{2}}$ be the spatial coordinates of two datasets, and f$_{1}$ and f$_{2}$ the corresponding spatial probability density functions. The kernel density estimates of f$_{1}$ and f$_{2}$ are

\begin{equation}
\hat{f_{1}}(x,H_{1}) = \frac{1}{n_{1}}\sum_{i=1}^{n_{1}} K_{H_{1}}(x-X_{i})
\end{equation}

\begin{equation}
\hat{f_{2}}(x,H_{2}) = \frac{1}{n_{2}}\sum_{i=1}^{n_{2}} K_{H_{2}}(x-X_{i})
\end{equation}

\noindent where $K$ is the kernel function with $K_{H_{l}}= |H_{l}|^{-1/2}K(H_{l}^{-1/2} x)$ \citep[][]{wandJones1993}, and $H_{l}$ is a bandwidth matrix, for $l=1,2$. To test the null hypothesis $H_{0}:f_{1}=f_{2}$, a discrepancy measure is introduced: $T=\int[f_1(x) -f_2(x)]^2\;dx$. Assuming that the null hypothesis holds, it can be shown that

\begin{equation}
\mu_T=\bigg[n_1^{-1}|H_1|^{-1/2} + n_2^{-1}|H_2|^{-1/2}\bigg]K(0),
\end{equation}

\begin{equation}
\sigma_T^2=3\Bigg[\int f(x)^3\;dx - \Big(\int f(x)^2\;dx\Big)^2\Bigg]
\end{equation}

\noindent and the $Z$-score is 

\begin{equation}
Z = \frac{T - \mu_{T}}{{\sigma_{T} \sqrt{{\frac{1}{n_{1}} + \frac{1}{n_{2}}}}}}
\end{equation}

The p-value is then computed from this $z$-score using standard software or tables. The complete automatic testing procedure is programmed in the $ks$ library in the open-source R programming language \cite{DUO07}.
\newpage

%\begin{table}[!ht]
%	\centering
%	\caption{Parameters describing our sample.\label{tableSample}}
%	\scriptsize
%	\begin{tabular}{ccccccccc}
%		\hline\hline
%		%  \hline
%		Yang ID & RA & DEC & Redshift & M$_{200}$ & R$_{200}$ & N$_{R200}$ & BAX & NORAS/REFLEX \\
%		\vspace{-0.2cm} \\
%		\hline

\begin{ThreePartTable}
\begin{center}
\begin{longtable}{ccccccccc}
 \caption{Group Sample \label{tab:long}} \\

\hline 
\hline

\multicolumn{1}{c}{Yang ID} & \multicolumn{1}{c}{RA} & \multicolumn{1}{c}{DEC} & \multicolumn{1}{c}{Redshift} & \multicolumn{1}{c}{Log M$_{200}$} & \multicolumn{1}{c}{R$_{200}$} & \multicolumn{1}{c}{N$_{\rm R200}$} & \multicolumn{1}{c}{BAX} & \multicolumn{1}{c}{NORAS/REFLEX}\\ 

\multicolumn{1}{c}{} & \multicolumn{1}{c}{$^{\circ}$} & \multicolumn{1}{c}{$^{\circ}$} & \multicolumn{1}{c}{} & \multicolumn{1}{c}{} & \multicolumn{1}{c}{Mpc} & \multicolumn{1}{c}{} & \multicolumn{1}{c}{} & \multicolumn{1}{c}{}\\ 

\hline 
\endfirsthead

\multicolumn{9}{c}%
 {{\tablename\ \thetable{} -- continued from previous page}} \\
\hline 

\multicolumn{1}{c}{Yang ID} & \multicolumn{1}{c}{RA} & \multicolumn{1}{c}{DEC} & \multicolumn{1}{c}{Redshift} & \multicolumn{1}{c}{Log M$_{200}$} & \multicolumn{1}{c}{R$_{200}$} & \multicolumn{1}{c}{N$_{\rm R200}$} & \multicolumn{1}{c}{BAX} & \multicolumn{1}{c}{NORAS/REFLEX}\\ 

\multicolumn{1}{c}{} & \multicolumn{1}{c}{$^{\circ}$} & \multicolumn{1}{c}{$^{\circ}$} & \multicolumn{1}{c}{} & \multicolumn{1}{c}{} & \multicolumn{1}{c}{Mpc} & \multicolumn{1}{c}{} & \multicolumn{1}{c}{} & \multicolumn{1}{c}{}\\ 

\hline 
\endhead

\hline \multicolumn{9}{r}{{Continued on next page}} \\ \hline
\endfoot

\hline \hline
\endlastfoot

000002 & 240.5602 &  16.1113 & 0.036 & 15.29 & 2.01 & 525 & Y & N \\
000004 & 247.1149 &  40.8317 & 0.030 & 14.66 & 1.24 & 217 & Y & Y \\
000005 & 247.1633 &  39.4674 & 0.030 & 14.81 & 1.40 & 277 & Y & Y \\
000006 & 167.6936 &  28.5374 & 0.033 & 14.60 & 1.18 & 167 & Y & N \\
000007 & 351.1194 &  14.6251 & 0.042 & 14.51 & 1.10 & 106 & Y & N \\
000008 & 239.5372 &  27.3133 & 0.090 & 15.10 & 1.71 & 183 & Y & Y \\
000009 & 241.5673 &  18.1483 & 0.038 & 14.48 & 1.08 & 143 & Y & N \\
000010 & 223.2712 &  16.7240 & 0.045 & 14.22 & 0.89 & 106 & Y & Y \\
000011 &  10.4693 &  -9.3997 & 0.056 & 14.96 & 1.55 & 156 & Y & Y \\
000012 &  14.2060 &  -0.7460 & 0.044 & 14.54 & 1.13 &  95 & N & Y \\
000014 & 228.1091 &   7.4642 & 0.045 & 14.49 & 1.09 &  96 & Y & N \\
000015 & 215.5357 &  48.4549 & 0.072 & 14.78 & 1.35 &  90 & N & N \\
000016 & 230.7995 &   8.6412 & 0.034 & 14.65 & 1.23 & 115 & Y & Y \\
000017 & 216.7229 &  16.7528 & 0.053 & 14.55 & 1.14 & 107 & Y & N \\
000018 & 168.0412 &  40.5435 & 0.075 & 14.39 & 0.99 &  41 & N & Y \\
000020 & 241.1981 &  17.5749 & 0.034 & 14.92 & 1.51 & 257 & Y & Y \\
000021 & 230.6338 &  27.7129 & 0.073 & 15.27 & 1.96 & 144 & Y & Y \\
000022 &  18.7854 &   0.3047 & 0.045 & 14.38 & 1.00 &  82 & Y & Y \\
000023 & 227.7485 &   6.0388 & 0.079 & 15.21 & 1.87 & 132 & N & N \\
000024 & 229.2044 &   7.0289 & 0.035 & 14.45 & 1.05 &  87 & Y & Y \\
000025 & 230.2953 &  30.6325 & 0.078 & 14.70 & 1.26 &  88 & Y & N \\
000026 & 258.2267 &  64.0514 & 0.081 & 15.14 & 1.77 & 150 & Y & Y \\
000027 & 169.1316 &  29.2563 & 0.047 & 14.29 & 0.93 &  76 & Y & N \\
000028 & 190.3042 &  18.5539 & 0.072 & 14.73 & 1.30 &  87 & Y & N \\
000029 & 159.7771 &   5.1123 & 0.069 & 14.59 & 1.16 &  74 & Y & N \\
000030 & 329.3637 &  -7.7877 & 0.058 & 14.45 & 1.05 &  86 & Y & Y \\
000031 & 234.9381 &  21.7405 & 0.041 & 14.58 & 1.17 & 109 & Y & N \\
000032 & 358.5345 & -10.3766 & 0.076 & 14.81 & 1.38 &  78 & Y & Y \\
000034 & 207.2288 &  26.6148 & 0.063 & 14.82 & 1.40 &  95 & Y & Y \\
000035 & 164.6340 &   1.6037 & 0.040 & 14.05 & 0.78 &  50 & Y & Y \\
000036 & 202.7342 &  -1.8838 & 0.086 & 14.82 & 1.38 &  77 & Y & Y \\
000037 & 241.5638 &  15.6789 & 0.040 & 15.17 & 1.83 & 221 & N & N \\
000038 & 173.6865 &  49.0815 & 0.033 & 14.34 & 0.97 &  98 & Y & N \\
000039 & 229.8875 &  20.8222 & 0.040 & 14.22 & 0.89 &  55 & N & N \\
000040 & 223.6396 &  18.6078 & 0.058 & 14.53 & 1.11 &  81 & Y & Y \\
000041 & 117.0449 &  18.5520 & 0.047 & 14.40 & 1.01 &  59 & Y & Y \\
000042 & 244.4189 &  35.0411 & 0.031 & 14.60 & 1.19 & 107 & Y & Y \\
000044 & 229.6038 &   4.5283 & 0.037 & 14.38 & 1.00 &  61 & N & N \\
000045 & 184.4023 &   3.6390 & 0.077 & 15.04 & 1.64 & 100 & N & Y \\
000046 & 127.1567 &  30.4313 & 0.050 & 14.69 & 1.26 &  83 & Y & Y \\
000047 & 241.3060 &  16.4602 & 0.044 & 15.15 & 1.80 & 232 & N & N \\
000048 &  14.0058 & -10.0165 & 0.055 & 14.50 & 1.09 &  72 & Y & N \\
000049 & 176.8442 &  55.7178 & 0.051 & 14.66 & 1.24 &  83 & Y & N \\
000050 & 214.3435 &   2.0353 & 0.054 & 14.50 & 1.09 &  69 & Y & N \\
000051 & 214.4332 &   8.1960 & 0.057 & 14.43 & 1.04 &  69 & N & N \\
000052 & 176.3217 &  33.3532 & 0.032 & 14.10 & 0.81 &  67 & N & N \\
000053 & 208.2782 &   5.1703 & 0.079 & 14.68 & 1.24 &  70 & Y & N \\
000054 & 203.9791 &  59.2357 & 0.071 & 14.86 & 1.43 & 108 & Y & Y \\
000057 & 180.0944 &  56.2254 & 0.064 & 14.62 & 1.20 &  69 & Y & N \\
000058 & 226.1477 &  28.4716 & 0.058 & 14.45 & 1.05 &  62 & Y & N \\
000061 & 173.2294 &  14.4326 & 0.081 & 14.83 & 1.40 &  61 & Y & N \\
000062 & 195.6601 &  -2.5423 & 0.083 & 14.62 & 1.18 &  57 & N & Y \\
000064 & 146.5710 &  43.1473 & 0.073 & 14.12 & 0.81 &  24 & N & N \\
000065 & 230.4156 &   7.7193 & 0.045 & 14.57 & 1.15 &  74 & Y & Y \\
000066 & 197.7928 &  39.2503 & 0.072 & 14.79 & 1.36 &  55 & Y & N \\
000068 & 205.4739 &  26.3809 & 0.075 & 14.46 & 1.05 &  39 & Y & Y \\
000069 & 146.7111 &  54.4917 & 0.047 & 14.45 & 1.06 &  66 & N & N \\
000070 & 167.1201 &  44.0939 & 0.059 & 14.38 & 1.00 &  66 & Y & N \\
000071 & 200.0644 &  33.1644 & 0.036 & 14.41 & 1.02 &  57 & Y & Y \\
000073 & 245.6562 &  37.9252 & 0.031 & 14.53 & 1.13 &  75 & N & N \\
000074 & 208.0007 &  46.3667 & 0.063 & 14.39 & 1.00 &  53 & Y & Y \\
000077 & 177.0543 &  54.6677 & 0.060 & 14.26 & 0.90 &  54 & N & N \\
000079 & 170.4099 &   2.8163 & 0.049 & 14.30 & 0.94 &  49 & Y & Y \\
000080 & 228.8030 &   4.3649 & 0.097 & 14.68 & 1.24 &  43 & N & N \\
000082 & 168.8992 &  54.5174 & 0.070 & 14.57 & 1.14 &  57 & Y & Y \\
000083 & 227.9499 &   4.5020 & 0.036 & 14.40 & 1.02 &  54 & N & N \\
000085 & 196.0038 &  19.2772 & 0.064 & 14.60 & 1.17 &  54 & Y & Y \\
000086 & 116.5758 &  18.3162 & 0.052 & 14.26 & 0.91 &  49 & N & N \\
000087 & 174.0717 &  55.0578 & 0.057 & 14.02 & 0.75 &  44 & Y & N \\
000088 & 233.1359 &   4.7342 & 0.038 & 14.04 & 0.77 &  38 & Y & N \\
000089 & 170.7403 &   1.0701 & 0.074 & 14.30 & 0.93 &  38 & Y & N \\
000090 & 122.4772 &  35.1596 & 0.083 & 14.67 & 1.23 &  41 & N & N \\
000093 & 186.5582 &  31.1117 & 0.060 & 13.92 & 0.70 &  26 & N & N \\
000094 & 179.2471 &   5.0477 & 0.075 & 14.59 & 1.16 &  53 & Y & N \\
000095 & 187.4236 &  11.7516 & 0.086 & 14.88 & 1.45 &  74 & Y & Y \\
000096 & 231.0865 &  20.7606 & 0.040 & 14.16 & 0.84 &  47 & N & N \\
000097 & 140.0518 &  54.8904 & 0.045 & 14.12 & 0.82 &  34 & N & N \\
000098 & 155.3920 &  23.9116 & 0.039 & 14.16 & 0.84 &  51 & N & N \\
000099 & 158.3152 &  56.7684 & 0.045 & 14.07 & 0.78 &  47 & Y & N \\
000100 & 183.6602 &  59.9266 & 0.060 & 14.22 & 0.88 &  47 & Y & N \\
000101 &  51.3858 &  -0.5800 & 0.037 & 13.81 & 0.64 &  22 & N & N \\
000102 & 165.2002 &  10.4066 & 0.036 & 14.65 & 1.23 &  65 & Y & N \\
000103 & 157.9400 &  40.1824 & 0.067 & 14.14 & 0.82 &  38 & Y & N \\
000104 & 236.2194 &  36.1064 & 0.066 & 14.70 & 1.27 &  67 & Y & N \\
000105 &   7.3223 &  -0.2126 & 0.060 & 14.20 & 0.87 &  33 & N & N \\
000106 & 172.3877 &  54.1117 & 0.069 & 14.41 & 1.02 &  51 & Y & N \\
000107 & 212.5518 &  54.9166 & 0.042 & 13.91 & 0.69 &  33 & N & N \\
000108 & 252.6651 &  23.5084 & 0.036 & 13.93 & 0.71 &  37 & N & N \\
000110 &  18.2482 &  15.4844 & 0.043 & 14.25 & 0.90 &  52 & Y & Y \\
000111 & 184.6299 &   5.2140 & 0.077 & 14.60 & 1.17 &  42 & Y & N \\
000112 & 238.8791 &  41.6157 & 0.034 & 14.20 & 0.87 &  38 & N & N \\
000114 & 176.4182 &  15.5056 & 0.068 & 14.27 & 0.91 &  37 & Y & N \\
000115 & 246.7696 &  14.1874 & 0.051 & 13.89 & 0.69 &  27 & N & N \\
000116 & 168.7743 &  25.8728 & 0.048 & 13.97 & 0.73 &  43 & N & N \\
000118 & 152.5112 &  54.4350 & 0.046 & 14.32 & 0.95 &  60 & N & N \\
000119 & 238.1031 &  27.7115 & 0.082 & 13.99 & 0.73 &  20 & N & N \\
000120 & 170.5107 &  34.3465 & 0.035 & 13.59 & 0.54 &  34 & Y & N \\
000121 & 194.6833 &  -1.7481 & 0.084 & 14.70 & 1.26 &  48 & Y & Y \\
000122 & 216.0279 &  26.2658 & 0.038 & 13.82 & 0.65 &  32 & N & Y \\
000125 &  24.3436 &  -9.1930 & 0.041 & 14.18 & 0.86 &  54 & Y & Y \\
000126 & 199.8197 &  -0.9386 & 0.083 & 14.62 & 1.19 &  37 & N & N \\
000127 &  17.4503 &  14.1071 & 0.060 & 14.54 & 1.12 &  51 & Y & Y \\
000128 & 240.3577 &  53.9616 & 0.065 & 13.98 & 0.73 &  24 & Y & Y \\
000129 & 175.3044 &   5.7204 & 0.097 & 14.27 & 0.90 &  23 & Y & N \\
000130 & 255.6457 &  33.5499 & 0.086 & 15.02 & 1.61 &  80 & Y & N \\
000132 & 207.3671 &  28.0318 & 0.076 & 14.59 & 1.16 &  43 & Y & Y \\
000135 & 207.9144 &   9.4982 & 0.065 & 14.51 & 1.09 &  43 & N & N \\
000136 & 222.1278 &  11.2865 & 0.052 & 14.37 & 0.99 &  49 & N & N \\
000137 & 150.6702 &  32.7208 & 0.051 & 14.09 & 0.80 &  46 & Y & Y \\
000140 & 155.4981 &  38.5215 & 0.054 & 14.54 & 1.13 &  63 & N & Y \\
000141 & 245.2292 &  29.8382 & 0.097 & 14.60 & 1.16 &  35 & Y & Y \\
000142 & 248.0229 &  13.6468 & 0.053 & 14.18 & 0.85 &  44 & N & N \\
000143 & 192.2485 &  -1.7069 & 0.087 & 14.97 & 1.55 &  65 & Y & N \\
000144 & 134.9522 &  39.3855 & 0.095 & 14.26 & 0.89 &  34 & Y & N \\
000147 & 228.0394 &   1.8068 & 0.039 & 14.09 & 0.80 &  21 & Y & Y \\
000148 & 257.4347 &  34.4547 & 0.085 & 15.03 & 1.63 &  76 & Y & Y \\
000149 & 196.4868 &   9.4838 & 0.055 & 14.04 & 0.77 &  27 & N & N \\
000151 & 248.8257 &  26.6174 & 0.070 & 13.81 & 0.64 &  21 & N & N \\
000153 & 239.6032 &  18.0424 & 0.046 & 14.38 & 1.00 &  52 & N & N \\
000154 & 161.7944 &  38.9640 & 0.036 & 14.25 & 0.91 &  33 & N & N \\
000156 & 156.2606 &  17.1350 & 0.045 & 13.80 & 0.64 &  22 & N & N \\
000157 & 232.3110 &  52.8639 & 0.073 & 14.27 & 0.91 &  34 & N & N \\
000158 & 163.5235 &  54.8282 & 0.072 & 14.39 & 1.00 &  42 & Y & N \\
000159 & 223.5304 &  54.2627 & 0.099 & 14.17 & 0.83 &  29 & N & N \\
000160 & 230.8448 &  31.0494 & 0.074 & 14.44 & 1.03 &  33 & Y & N \\
000164 & 170.6277 &  34.0755 & 0.043 & 13.87 & 0.68 &  36 & N & N \\
000166 & 138.2210 &  47.7280 & 0.051 & 14.14 & 0.83 &  32 & Y & N \\
000168 & 205.9135 &  55.6228 & 0.068 & 13.93 & 0.70 &  34 & Y & N \\
000170 & 232.3015 &   7.5732 & 0.043 & 14.03 & 0.76 &  29 & N & N \\
000171 & 173.7788 &  21.1512 & 0.064 & 13.85 & 0.66 &  22 & N & N \\
000172 &  10.7605 &  15.2429 & 0.079 & 14.34 & 0.96 &  32 & N & N \\
000173 &  27.3010 &  14.0424 & 0.070 & 13.88 & 0.68 &  22 & Y & N \\
000174 & 134.5701 &  38.5280 & 0.093 & 14.23 & 0.88 &  32 & N & N \\
000175 & 151.0412 &  54.6353 & 0.047 & 14.39 & 1.00 &  43 & N & N \\
000176 & 227.7767 &   5.2948 & 0.080 & 14.76 & 1.32 &  37 & N & N \\
000179 & 248.3581 &  11.8340 & 0.052 & 14.23 & 0.89 &  37 & N & N \\
000180 & 260.6725 &  30.8284 & 0.046 & 14.26 & 0.91 &  39 & Y & Y \\
000181 & 227.4433 &   8.8004 & 0.080 & 14.47 & 1.06 &  29 & N & N \\
000182 & 233.3925 &  31.0837 & 0.067 & 14.25 & 0.89 &  30 & Y & N \\
000183 &  30.5650 &  -1.0403 & 0.042 & 14.04 & 0.77 &  28 & Y & Y \\
000184 & 117.3782 &  52.0970 & 0.069 & 14.37 & 0.98 &  41 & Y & N \\
000186 & 203.1493 &  32.5997 & 0.036 & 14.05 & 0.78 &  40 & N & N \\
000187 & 188.8342 &   1.8310 & 0.080 & 14.23 & 0.88 &  23 & Y & N \\
000188 &  59.3620 &  -5.3835 & 0.066 & 13.87 & 0.67 &  20 & N & N \\
000189 & 182.5888 &   5.3469 & 0.077 & 14.61 & 1.18 &  47 & Y & Y \\
000190 & 254.1129 &  39.2791 & 0.062 & 14.07 & 0.78 &  31 & N & N \\
000192 & 218.3383 &  52.8728 & 0.045 & 14.05 & 0.78 &  36 & N & N \\
000193 & 156.7968 &  11.0160 & 0.032 & 13.37 & 0.46 &  27 & Y & N \\
000194 & 239.1577 &  25.8093 & 0.072 & 14.31 & 0.94 &  22 & N & N \\
000196 & 206.1913 &  29.7351 & 0.062 & 14.09 & 0.80 &  30 & Y & N \\
000198 & 136.9208 &  52.0922 & 0.061 & 14.12 & 0.81 &  36 & N & N \\
000199 & 122.6209 &  42.2998 & 0.064 & 14.17 & 0.85 &  31 & Y & N \\
000201 & 129.8160 &  28.7763 & 0.080 & 14.43 & 1.03 &  30 & Y & N \\
000203 & 186.8957 &   8.8389 & 0.089 & 14.87 & 1.44 &  68 & Y & Y \\
000204 & 229.9829 &  25.7578 & 0.033 & 14.06 & 0.78 &  36 & N & N \\
000205 & 225.5257 &  21.2966 & 0.062 & 14.33 & 0.96 &  31 & N & N \\
000206 & 135.5553 &  20.6192 & 0.082 & 14.29 & 0.92 &  29 & N & N \\
000207 & 240.8769 &  14.7888 & 0.036 & 14.25 & 0.90 &  49 & N & N \\
000208 & 168.1048 &  57.0575 & 0.047 & 14.09 & 0.80 &  33 & N & N \\
000209 & 209.9132 &  28.5390 & 0.063 & 14.17 & 0.85 &  29 & N & N \\
000210 & 234.9841 &  17.8575 & 0.090 & 14.65 & 1.21 &  41 & Y & N \\
000212 & 162.9201 &  55.3390 & 0.074 & 13.91 & 0.69 &  29 & N & N \\
000213 & 180.7591 &  54.7744 & 0.050 & 14.06 & 0.78 &  31 & N & N \\
000216 & 243.2548 &  30.9166 & 0.050 & 14.51 & 1.10 &  43 & N & N \\
000221 & 183.2659 &  59.2576 & 0.096 & 13.85 & 0.65 &  21 & Y & N \\
000222 & 242.8492 &  36.9950 & 0.067 & 14.10 & 0.80 &  23 & N & N \\
000223 & 206.6898 &  45.7150 & 0.065 & 13.97 & 0.72 &  28 & N & N \\
000224 & 133.5641 &  29.0514 & 0.084 & 14.58 & 1.15 &  33 & N & N \\
000228 & 227.8607 &  -0.1083 & 0.091 & 14.22 & 0.87 &  25 & Y & N \\
000230 & 128.9867 &  38.5113 & 0.057 & 14.07 & 0.79 &  31 & N & N \\
000231 & 191.7995 &  55.0208 & 0.083 & 14.42 & 1.02 &  25 & Y & N \\
000232 & 230.2374 &  48.6807 & 0.074 & 14.53 & 1.11 &  34 & N & Y \\
000233 & 167.9791 &  39.6287 & 0.076 & 14.17 & 0.84 &  22 & N & N \\
000234 &  41.6991 &  -0.5708 & 0.043 & 13.49 & 0.50 &  21 & N & N \\
000235 & 233.8795 &  25.2250 & 0.034 & 13.53 & 0.52 &  23 & N & N \\
000236 & 205.5982 &  29.8260 & 0.044 & 14.26 & 0.91 &  35 & N & N \\
000238 & 120.9997 &  10.0498 & 0.034 & 13.64 & 0.57 &  24 & N & N \\
000239 & 119.1450 &  45.7419 & 0.052 & 14.20 & 0.87 &  29 & N & N \\
000240 & 191.2993 &   1.7884 & 0.048 & 13.85 & 0.66 &  25 & N & N \\
000241 & 354.4769 &  15.8294 & 0.067 & 14.20 & 0.87 &  29 & N & N \\
000242 & 136.9767 &  49.6365 & 0.035 & 13.94 & 0.71 &  31 & N & N \\
000243 & 162.4875 &   0.3461 & 0.039 & 13.94 & 0.71 &  26 & N & N \\
000245 & 130.6866 &  36.0876 & 0.054 & 14.18 & 0.85 &  30 & N & Y \\
000246 & 132.1697 &   9.0540 & 0.064 & 14.13 & 0.82 &  27 & N & N \\
000247 & 198.4603 &  38.9207 & 0.070 & 14.36 & 0.98 &  31 & N & N \\
000250 & 189.7137 &  17.1924 & 0.071 & 13.92 & 0.70 &  20 & N & N \\
000253 & 173.0947 &  56.2955 & 0.050 & 14.15 & 0.84 &  32 & Y & Y \\
000254 & 212.1701 &  55.5179 & 0.074 & 14.27 & 0.91 &  25 & N & N \\
000255 & 233.2867 &  27.9916 & 0.073 & 14.46 & 1.05 &  32 & Y & N \\
000256 & 236.8236 &  27.9889 & 0.076 & 14.21 & 0.87 &  23 & N & N \\
000258 & 247.6049 &  24.5827 & 0.064 & 14.20 & 0.86 &  32 & Y & N \\
000259 & 155.8174 &  12.9092 & 0.032 & 13.81 & 0.65 &  25 & Y & Y \\
000260 & 173.9414 &  13.6814 & 0.080 & 14.48 & 1.06 &  31 & N & N \\
000263 & 207.0932 &  25.6780 & 0.052 & 14.10 & 0.80 &  28 & N & N \\
000265 & 209.1915 &  23.1310 & 0.063 & 14.36 & 0.98 &  38 & N & N \\
000267 & 195.3458 &  -3.4901 & 0.085 & 14.21 & 0.86 &  21 & N & N \\
000268 & 177.5320 &   5.6780 & 0.075 & 14.20 & 0.86 &  21 & N & N \\
000269 & 192.9966 &   4.5747 & 0.065 & 13.87 & 0.67 &  26 & Y & N \\
000272 & 177.4340 &  12.3262 & 0.083 & 14.15 & 0.83 &  25 & N & N \\
000275 & 244.7847 &  24.2078 & 0.066 & 14.07 & 0.78 &  24 & N & N \\
000276 & 233.8128 &  27.3153 & 0.032 & 14.10 & 0.81 &  30 & N & N \\
000278 & 209.6037 &  32.6541 & 0.049 & 14.00 & 0.74 &  35 & N & N \\
000279 & 219.3622 &  24.8378 & 0.088 & 14.20 & 0.86 &  21 & Y & N \\
000280 &  19.7378 &  -1.0177 & 0.045 & 14.10 & 0.81 &  24 & Y & N \\
000281 & 126.6305 &  17.4708 & 0.089 & 13.93 & 0.70 &  21 & N & N \\
000282 & 212.7563 &  19.1441 & 0.056 & 13.94 & 0.71 &  25 & N & N \\
000283 & 222.1024 &  18.3360 & 0.040 & 14.16 & 0.84 &  37 & N & N \\
000287 & 190.5543 &  -2.1150 & 0.083 & 14.21 & 0.87 &  25 & N & N \\
000289 & 217.4548 &  53.9933 & 0.043 & 13.77 & 0.63 &  24 & N & N \\
000290 & 197.1580 &  43.5216 & 0.036 & 13.99 & 0.74 &  30 & N & N \\
000292 & 161.1207 &  14.1048 & 0.033 & 13.98 & 0.73 &  29 & N & N \\
000294 & 167.4929 &  21.8103 & 0.032 & 13.76 & 0.62 &  24 & Y & Y \\
000295 & 131.8533 &  53.8186 & 0.045 & 14.12 & 0.82 &  30 & N & N \\
000297 & 211.5987 &   6.2928 & 0.085 & 14.30 & 0.93 &  23 & N & N \\
000298 & 156.2628 &  47.8127 & 0.063 & 14.27 & 0.92 &  31 & N & N \\
000300 & 186.2859 &  61.4903 & 0.070 & 13.83 & 0.65 &  23 & Y & N \\
000301 & 172.4102 &  55.4223 & 0.068 & 14.07 & 0.78 &  23 & N & N \\
000302 & 158.0672 &  40.2840 & 0.078 & 14.25 & 0.89 &  26 & N & N \\
000303 & 236.0939 &  34.6402 & 0.070 & 14.44 & 1.04 &  22 & N & N \\
000304 & 185.5021 &  13.7596 & 0.081 & 14.42 & 1.02 &  24 & Y & N \\
000305 & 217.4873 &   7.2445 & 0.054 & 14.19 & 0.86 &  22 & N & N \\
000306 & 230.1077 &  33.4153 & 0.082 & 14.20 & 0.86 &  23 & N & N \\
000308 & 235.7148 &   8.2405 & 0.041 & 13.99 & 0.74 &  25 & N & N \\
000309 & 230.7480 &  28.5960 & 0.084 & 14.35 & 0.96 &  22 & N & N \\
000310 & 160.3120 &  33.8973 & 0.083 & 14.11 & 0.80 &  24 & N & N \\
000311 & 242.9673 &  20.9786 & 0.086 & 14.35 & 0.97 &  24 & N & N \\
000313 & 137.1075 &  16.0310 & 0.072 & 14.30 & 0.93 &  31 & N & N \\
000314 & 168.4237 &   2.4990 & 0.074 & 14.05 & 0.77 &  30 & N & N \\
000316 & 157.1190 &   3.7424 & 0.073 & 14.28 & 0.92 &  26 & N & N \\
000319 & 246.9982 &  42.6871 & 0.032 & 13.86 & 0.67 &  26 & N & Y \\
000322 & 159.2213 &  50.1229 & 0.045 & 14.15 & 0.83 &  32 & N & N \\
000327 & 225.8592 &   7.9980 & 0.089 & 14.62 & 1.19 &  32 & N & N \\
000328 & 253.5900 &  23.5416 & 0.057 & 14.14 & 0.82 &  31 & Y & Y \\
000330 & 216.1921 &  29.5507 & 0.054 & 13.98 & 0.73 &  25 & N & N \\
000331 & 118.5921 &  14.6519 & 0.050 & 13.91 & 0.69 &  28 & N & N \\
000333 & 187.7690 &  28.8527 & 0.062 & 14.06 & 0.78 &  26 & N & N \\
000336 & 217.1839 &  17.4129 & 0.052 & 14.07 & 0.78 &  26 & N & N \\
000338 & 155.6189 &  15.7453 & 0.045 & 13.87 & 0.67 &  23 & N & N \\
000340 & 213.6193 &   1.7645 & 0.054 & 13.99 & 0.74 &  25 & N & N \\
000341 & 157.5611 &   4.0191 & 0.067 & 13.76 & 0.62 &  22 & N & N \\
000342 & 122.1481 &  38.9195 & 0.041 & 13.75 & 0.62 &  27 & N & N \\
000349 & 175.7910 &  26.4327 & 0.031 & 13.78 & 0.63 &  21 & N & N \\
000351 & 193.6675 &  18.9850 & 0.064 & 14.02 & 0.75 &  25 & Y & N \\
000354 & 244.7571 &  50.4838 & 0.056 & 14.35 & 0.97 &  24 & N & N \\
000355 & 189.0735 &  16.5470 & 0.069 & 14.56 & 1.14 &  34 & N & N \\
000358 & 205.5057 &   2.1496 & 0.077 & 14.72 & 1.29 &  53 & Y & Y \\
000368 & 131.6204 &  29.6226 & 0.071 & 14.24 & 0.89 &  20 & N & N \\
000371 & 181.3786 &  43.1855 & 0.053 & 14.12 & 0.82 &  24 & N & N \\
000373 & 226.1811 &  35.9573 & 0.048 & 14.06 & 0.78 &  25 & N & N \\
000374 & 245.7299 &  23.8919 & 0.063 & 14.10 & 0.80 &  31 & N & N \\
000375 & 145.1331 &  66.6577 & 0.070 & 14.33 & 0.96 &  20 & Y & N \\
000377 & 203.9595 &  35.9609 & 0.059 & 14.22 & 0.88 &  27 & N & N \\
000381 & 249.2461 &  44.3289 & 0.031 & 13.80 & 0.64 &  21 & N & N \\
000388 & 122.7397 &  35.9724 & 0.082 & 14.48 & 1.06 &  31 & Y & N \\
000389 & 224.5938 &  47.4751 & 0.086 & 14.45 & 1.04 &  26 & N & N \\
000390 & 231.4995 &  48.4927 & 0.037 & 13.51 & 0.51 &  21 & N & N \\
000394 & 199.1440 &   7.0911 & 0.049 & 14.11 & 0.81 &  25 & N & N \\
000396 & 232.9112 &   7.3305 & 0.034 & 13.73 & 0.61 &  20 & N & N \\
000397 & 201.2498 &   8.3651 & 0.052 & 13.82 & 0.65 &  21 & N & N \\
000408 & 189.8714 &  16.5063 & 0.070 & 14.32 & 0.95 &  24 & N & N \\
000409 & 192.5124 &  -1.5625 & 0.082 & 14.84 & 1.40 &  50 & N & N \\
000412 & 116.4103 &  33.9561 & 0.062 & 14.29 & 0.93 &  29 & N & N \\
000413 & 255.6881 &  33.4648 & 0.092 & 15.71 & 2.73 & 174 & N & N \\
000414 & 117.4209 &  29.4346 & 0.063 & 14.24 & 0.89 &  25 & N & N \\
000416 & 174.8016 &  55.6594 & 0.062 & 14.29 & 0.92 &  26 & N & N \\
000418 & 181.9639 &  14.9781 & 0.079 & 14.50 & 1.09 &  34 & Y & N \\
000420 & 242.9955 &  29.8385 & 0.050 & 14.71 & 1.29 &  45 & N & N \\
000423 & 195.2958 &  39.8310 & 0.036 & 13.99 & 0.74 &  24 & N & N \\
000424 & 164.3707 &  37.6233 & 0.035 & 13.68 & 0.58 &  21 & N & N \\
000427 & 186.3138 &  32.1500 & 0.059 & 14.35 & 0.97 &  38 & Y & Y \\
000432 & 168.2233 &   2.5009 & 0.078 & 15.11 & 1.73 &  66 & Y & N \\
000434 & 243.3893 &  49.0820 & 0.056 & 14.33 & 0.95 &  47 & N & N \\
000435 & 197.2963 &  -1.6222 & 0.083 & 14.54 & 1.11 &  23 & N & Y \\
000436 & 126.5674 &  40.9862 & 0.057 & 14.02 & 0.75 &  21 & N & N \\
000439 & 229.7450 &   4.3358 & 0.047 & 14.06 & 0.78 &  24 & N & N \\
000440 & 215.4074 &  49.5070 & 0.072 & 14.51 & 1.10 &  24 & N & Y \\
000441 & 229.9992 &  32.8171 & 0.080 & 14.10 & 0.80 &  23 & N & N \\
000445 & 171.6007 &  35.3344 & 0.034 & 13.82 & 0.65 &  20 & N & N \\
000447 & 250.3106 &  13.3536 & 0.051 & 13.96 & 0.72 &  25 & N & N \\
000452 & 331.7979 &  -7.5310 & 0.061 & 14.17 & 0.85 &  28 & N & N \\
000458 & 145.9013 &  39.3971 & 0.041 & 14.01 & 0.75 &  26 & N & N \\
000460 & 131.2250 &  27.7190 & 0.084 & 14.12 & 0.81 &  20 & Y & N \\
000467 & 237.3127 &  25.6057 & 0.071 & 14.12 & 0.81 &  22 & N & N \\
000471 & 229.4987 &  26.8780 & 0.085 & 14.67 & 1.23 &  31 & N & N \\
000480 & 167.7640 &   1.1049 & 0.097 & 13.92 & 0.69 &  21 & Y & N \\
000496 & 227.3806 &   7.5358 & 0.077 & 14.59 & 1.16 &  32 & N & N \\
000501 & 202.3909 &  37.4487 & 0.057 & 13.82 & 0.65 &  20 & Y & N \\
000503 & 197.6967 &  34.4158 & 0.037 & 14.32 & 0.95 &  27 & N & N \\
000509 & 139.8781 &  55.5739 & 0.048 & 14.00 & 0.74 &  23 & N & N \\
000511 & 237.7877 &  53.4282 & 0.065 & 13.71 & 0.59 &  22 & N & N \\
000513 & 250.3571 &  40.1635 & 0.032 & 14.46 & 1.07 &  32 & N & N \\
000526 & 237.2284 &   8.8457 & 0.072 & 14.67 & 1.24 &  30 & N & N \\
000529 & 243.8981 &  19.4815 & 0.031 & 13.70 & 0.60 &  20 & Y & Y \\
000530 & 207.1275 &  32.2499 & 0.083 & 14.24 & 0.89 &  21 & Y & N \\
000532 & 179.4821 &  33.7051 & 0.080 & 14.30 & 0.93 &  21 & N & Y \\
000540 & 227.1193 &  -0.2580 & 0.090 & 14.65 & 1.21 &  37 & Y & N \\
000542 & 243.6301 &  49.1877 & 0.059 & 14.30 & 0.93 &  46 & Y & N \\
000554 & 132.5320 &  32.7757 & 0.066 & 13.90 & 0.69 &  20 & N & N \\
000557 & 167.3308 &  41.5666 & 0.077 & 14.25 & 0.90 &  25 & Y & Y \\
000559 & 218.2858 &   9.6789 & 0.086 & 14.23 & 0.88 &  20 & N & N \\
000562 & 340.8330 &   0.4133 & 0.058 & 13.97 & 0.73 &  20 & N & N \\
000564 & 194.9210 &  38.8014 & 0.034 & 13.35 & 0.45 &  22 & N & N \\
000567 & 226.2659 &  26.0287 & 0.054 & 14.04 & 0.77 &  26 & Y & N \\
000574 & 166.8872 &  15.8625 & 0.093 & 14.60 & 1.17 &  29 & N & N \\
000577 & 140.9969 &  13.1923 & 0.080 & 14.40 & 1.00 &  22 & N & N \\
000586 & 349.7253 & -10.2361 & 0.032 & 13.78 & 0.63 &  21 & N & N \\
000595 & 188.7150 &  50.7791 & 0.040 & 13.85 & 0.66 &  22 & N & N \\
000599 & 215.9747 &  40.2588 & 0.082 & 14.35 & 0.96 &  21 & Y & Y \\
000608 &   4.1949 &  -0.4609 & 0.065 & 14.26 & 0.91 &  21 & N & N \\
000616 & 172.1451 &  26.8291 & 0.053 & 14.15 & 0.84 &  28 & N & N \\
000621 & 215.2345 &  17.6660 & 0.051 & 14.21 & 0.87 &  35 & Y & N \\
000624 & 129.0212 &  52.7156 & 0.044 & 14.16 & 0.85 &  20 & N & N \\
000625 & 242.3892 &  53.0522 & 0.063 & 14.30 & 0.93 &  22 & N & N \\
000628 & 325.7397 &  -6.8996 & 0.054 & 14.34 & 0.96 &  40 & Y & N \\
000630 & 167.5969 &   4.8471 & 0.030 & 13.70 & 0.60 &  21 & N & N \\
000631 & 119.6350 &  37.7935 & 0.041 & 13.89 & 0.69 &  28 & N & N \\
000635 & 203.2348 &  60.1190 & 0.072 & 14.54 & 1.12 &  30 & N & N \\
000664 & 208.4432 &  33.1695 & 0.050 & 14.26 & 0.91 &  29 & N & N \\
000665 & 205.9073 &  30.0654 & 0.040 & 15.00 & 1.61 &  83 & N & N \\
 
\end{longtable}
\begin{tablenotes}
	\small
	\item Notes: Column (1) - identification number in Yang catalog; Columns (2) and (3) - right ascension and 
	declination in J2000; Column (4) mean redshift of the group estimated with the shift gapper technique;
	Columns (5) and (6) - dynamical mass (in M$_{\odot}$) and virial radius (in Mpc); Column (7) - number
	of member galaxies within R$_{200}$ down to the limiting magnitude of the catalog; Columns (8) and (9)
	indication of counterpart in BAX (http://bax.ast.obs-mip.fr/) or NORAS, \citep{Bohringeretal2000}, and
	REFLEX, \citep{Bohringeretal2004}
\end{tablenotes}
\end{center}
\end{ThreePartTable}

\begin{table}[!h]
	\centering
	\caption{  Performance of MCLUST and HD based on simulated data.}
	\label{TablecomparisonMCLUSTHD}
	\vskip 0.2 truecm
	\scriptsize
	\begin{tabular}{ccccccccccccc}
		\hline\hline
		% \multicolumn{number of columns}{position}{text}
		& \multicolumn{2}{c}{$0.5 \leq \pi \leq 0.9$} &  
		\multicolumn{2}{c}{$0.5 \leq \pi \leq 0.9$} & 
		\multicolumn{2}{c}{$\pi = 0.5$} &  
		\multicolumn{2}{c}{$\pi = 0.5$} & 
		\multicolumn{2}{c}{$\pi = 0.9$} & 
		\multicolumn{2}{c}{$\pi = 0.9$} \\
		& \multicolumn{2}{c}{$\sigma_1 = \sigma_2$}  & 
		\multicolumn{2}{c}{$\sigma_1 = 2 \sigma_2$}   &
		\multicolumn{2}{c}{$\sigma_1 = \sigma_2$}   &
		\multicolumn{2}{c}{$\sigma_1 = 2 \sigma_2$}   &
		\multicolumn{2}{c}{$\sigma_1 = \sigma_2$}   &
		\multicolumn{2}{c}{$\sigma_1 = 2 \sigma_2$}  \\
		%  \hline
		N$_{\rm points}$ &  HD &  MCLUST &  HD &  MCLUST &  HD &  MCLUST &  HD &  MCLUST &  HD &  MCLUST &  HD &  MCLUST \\
		\vspace{-0.2cm} \\
		\hline
		20 & 4.9 & 5.2 & 8.6 & 9.5 & 5.2 & 4.8 & 6.6 & 6.6 & 5.4 & 6.3 & 9.6 & 11.5 \\
		30 & 4.3 & 4.7 & 7.5 & 8.1 & 4.4 & 4.4 & 5.3 & 5.9 & 4.6 & 5.1 & 8.5 & 9.5 \\
		40 & 3.8 & 4.3 & 6.9 & 7.6 & 3.7 & 4.1 & 3.9 & 5.4 & 4.1 & 4.6 & 7.8 & 8.7 \\
		50 & 3.6 & 4.0 & 6.6 & 7.3 & 3.6 & 3.9 & 3.6 & 4.8 & 3.8 & 4.4 & 7.2 & 8.1 \\
		100 & 2.9 & 3.3 & 5.4 & 6.3 & 2.8 & 3.3 & 2.0 & 3.4 & 3.1 & 3.6 & 6.1 & 6.9 \\
		200 & 2.5 & 2.9 & 4.6 & 5.6 & 2.4 & 2.9 & 1.4 & 2.0 & 2.7 & 3.0 & 5.4 & 6.1 \\
		400 & 2.1 & 2.5 & 2.7 & 5.0 & 1.9 & 2.5 & 0.8 & 1.2 & 2.4 & 2.6 & 4.5 & 5.4 \\
		\hline  
	\end{tabular}
	\small
	\begin{tablenotes}
      \item {Notes: The quantities listed represent $\delta = \mid \mu_{1}/\sigma_{1} -  \mu_{2}/\sigma_{2}\mid$, where $\mu$ and $\sigma$ are the mean and standard deviation of the two gaussians; $\pi$ is the proportion of samples in both groups; and N$_{points}$ is the number of points sampling the gaussians.}
    \end{tablenotes}
  
\end{table}

\begin{table}[!ht]
	\centering
	\caption{Comparison of the PPS of G and NG systems in the bright and faint regimes, using the KDE test.}\label{tablePPSkde}
	\scriptsize
	\begin{tabular}{ccc}
		\hline\hline
		%  \hline
		Sample &  N$_{cases}$ &  Are they Similar ? \\
		~~ & ~~~ & (Statistically) \\
		\vspace{-0.2cm} \\
		\hline
		
		GB X GF & 0 & Y \\
		GB X NGB & 1 & Y \\
		% GB x NGF & 1000 & N \\
		% GF X NGB & 7 & Y \\
		GF X NGF & 723 &  N \\
		NGB X NGF & 951 & N \\
		\hline  
	\end{tabular}
\end{table}

\begin{table}[!ht]
	\centering
	\caption{Comparative analysis of the different Ad Hoc definition of regions of the PPS.\label{tablePPSanalysis}}
	\scriptsize
	\begin{tabular}{cccccc}
		\hline\hline
		Environment/Sample & Type & Fraction & Log M$_{stellar}$ &  Age & [Z/H] \\
		\hline
		{\bf GAUSSIAN/BRIGHT} & ~~ & ~~ & ~~ &  Gyr & ~~ \\
		\hline
		Inner Region & LV & 65\% & 11.04 & 8.25  & 0.044 \\
		~~ & HV & 23\% & 10.96 & 7.56  & 0.024 \\
		~~ & p-value & - & 0.0008 & 0.0144  & 0.0013 \\
		\hline
		
		Intermediate Region & LV & 64\% & 10.99 & 7.03  & 0.032 \\
		~~ & HV & 21\% & 10.96 & 5.88  & -0.006 \\
		~~ & p-value & - & 0.1464 & $< 10^{-4}$  &  $< 10^{-4}$ \\
		\hline
		
		Outer Region & LV & 68\% & 10.96 & 6.20  & 0.009 \\
		~~ & HV & 16\% & 10.97 & 5.60  & -0.023 \\
		~~ & p-value & - & 0.9998 & 0.0827  & 0.1267 \\
		\hline
		\vspace{0.2cm}
		{\bf GAUSSIAN/FAINT} & ~~ & ~~ & ~~ &  ~~ & ~~ \\
		\hline
		Inner Region & LV & 65\% & 10.19 & 5.91  & -0.059 \\
		~~ & HV & 24\% & 10.10 & 5.67  & -0.076 \\
		~~ & p-value & - & 0.2423 & 0.4642 & 0.66 \\
		\hline
		
		Intermediate Region & LV & 55\% & 10.12 & 3.82  & -0.214 \\
		~~ & HV & 25\% & 10.05 & 4.65  & -0.144 \\
		~~ & p-value & - & 0.2336 & 0.3874  &  0.0878 \\
		\hline
		
		Outer Region & LV & 67\% & 10.05 & 2.73  & -0.285 \\
		~~ & HV & 16\% & 9.92 & 2.70  & -0.257 \\
		~~ & p-value & - & 0.2805 & 0.9871  & 0.9991 \\
		\hline

		\vspace{0.2cm}
		{\bf NON-GAUSSIAN/BRIGHT} & ~~ & ~~ & ~~ &  ~~ & ~~ \\
		\hline
		Inner Region & LV & 65\% & 11.00 & 7.35  & -0.042 \\
		~~ & HV & 19\% & 11.03 & 7.04  & -0.022 \\
		~~ & p-value & - & 0.3520 & 0.6042 & 0.0613 \\
		\hline
		
		Intermediate Region & LV & 59\% & 10.98 & 6.53  & 0.033 \\
		~~ & HV & 28\% & 11.02 & 6.22  & 0.022 \\
		~~ & p-value & - & 0.2945 & 0.2939  &  0.4841 \\
		\hline
		
		Outer Region & LV & 69\% & 10.97 & 6.29  & 0.017 \\
		~~ & HV & 18\% & 11.04 & 6.82  & 0.024 \\
		~~ & p-value & - & 0.0985 & 0.1185  & 0.4632 \\
		\hline

		\vspace{0.2cm}
		
		{\bf NON-GAUSSIAN/FAINT} & ~~ & ~~ & ~~ &  ~~ & ~~ \\
		\hline
		Inner Region & LV & 64\% & 10.11 & 5.56  & -0.077 \\
		~~ & HV & 21\% & 10.02 & 4.70  & -0.111 \\
		~~ & p-value & - & 0.2123 & 0.0058 & 0.3000 \\
		\hline
		
		Intermediate Region & LV & 57\% & 9.98 & 4.00  & -0.146 \\
		~~ & HV & 29\% & 9.99 & 3.10  & -0.225 \\
		~~ & p-value & - & 0.9325 & 0.0429  &  0.0085 \\
		\hline
		
		Outer Region & LV & 60\% & 10.05 & 3.88  & -0.171 \\
		~~ & HV & 24\% & 10.03 & 3.56  & -0.193 \\
		~~ & p-value & - & 0.8904 & 0.6795  & 0.8593 \\
		\hline           
	\end{tabular}
	
	\begin{tablenotes}
	\normalsize
      \item {Notes: Mean values of Log M$_{stellar}$, Age and [Z/H] in different locations of the PPS. The nomenclature is: LV $\equiv \hspace{5pt} \mid \Delta V/\sigma \mid < 0.5$; HV  $\equiv \hspace{5pt} \mid \Delta V/\sigma \mid > 0.5$; Inner Region $\equiv \rm R/R_{200} < 0.5$; Intermediate Region $\equiv 0.5 < \rm R/R_{200} < 1.0$ and Outer Region $\equiv \rm R/R_{200} > 0.5$.}
    \end{tablenotes}
\end{table}

\begin{table}
	\centering
	\caption{p-values for the permutation test (in parenthesis, below, p-values for KDE test) when comparing  VIR, BS and INF regions for a given environment, G or NG systems.\label{tablepvalues}}
	\vskip 0.2 truecm
	\scriptsize
	\begin{tabular}{cccccc}
		\hline\hline
		%  \hline
		Region of Phase Space & Gaussianity & Mag Regime & Age &  [Z/H] & M$_{stellar}$ \\
		\vspace{-0.2cm} \\
		\hline
		VIR X INF & G & Bright & 0.002 & 0.002  & 0.002 \\
		~~~~ & ~~~~ & ~~~~~ & ($<$ 0.0001) & ($<$ 0.0001) & (0.0003) \\
		VIR X BS & G & Bright & 0.002 & 0.002  & 0.002 \\
		~~~~ & ~~~~ & ~~~~~ & ($<$ 0.0001) & ($<$ 0.0001) & ($<$ 0.0001) \\
		INF X BS & G & Bright & 0.104 & 0.012  & 0.838 \\
		~~~~ & ~~~~ & ~~~~~ & (0.062) & (0.033) & (0.944) \\
		
		\hline  
		
		VIR X INF & G & Faint & 0.002 & 0.002 & 0.002 \\
		~~~~ & ~~~~ & ~~~~~ & ($<$ 0.0001) & ($<$ 0.0001) & (0.0008) \\
		VIR X BS & G & Faint & 0.002 & 0.004 & 0.068\\
		~~~~ & ~~~~ & ~~~~~ & ($<$ 0.0001) & (0.0002) & (0.074) \\
		INF X BS & G & Faint & 0.192 & 0.336 & 0.116 \\
		~~~~ & ~~~~ & ~~~~~ & (0.372) & (0.499) & (0.101) \\
		
		\hline  
		
		VIR X INF & NG & Bright & 0.044 & 0.076 & 0.982 \\
		~~~~ & ~~~~ & ~~~~~ & (0.134) & (0.017) & (0.594) \\
		VIR X BS & NG & Bright & 0.044 & 0.004 & 0.120 \\
		~~~~ & ~~~~ & ~~~~~ & (0.014) & (0.001) & (0.072) \\
		INF X BS & NG & Bright & 0.988 & 0.694 & 0.224 \\
		~~~~ & ~~~~ & ~~~~~ & (0.681) & (0.728) & (0.064) \\
		
		\hline  
		
		VIR X INF & NG & Faint & 0.044 & 0.016 & 0.110 \\
		~~~~ & ~~~~ & ~~~~~ & (0.044) & (0.010) & (0.067) \\
		VIR X BS & NG & Faint & 0.010 & 0.002 & 0.316 \\
		~~~~ & ~~~~ & ~~~~~ & (0.010) & (0.006) & (0.202) \\
		INF X BS & NG & Faint & 0.760 & 0.564 & 0.456 \\
		~~~~ & ~~~~ & ~~~~~ & (0.938) & (0.388) & (0.335) \\
		
		\hline  
	\end{tabular}
	\begin{tablenotes}
	\normalsize
      \item {Notes. Positions in the PPS defined as in \cite{MMR11}, which is based on the cosmological simulation described in \cite{BorganiEtal2004}. VIR for Virial, BS for Backsplash and INF for Infall. As for the magnitude regime,  Bright means $\rm M_{\rm r} \leq -20.55$, which is the limiting absolute magnitude corresponding to the spectroscopic completeness of SDSS-DR7 at z = 0.1, Faint means $-20.55 < \rm M_{\rm r} \leq -18.40$, where the limiting absolute magnitude corresponds to the spectroscopic completeness of SDSS-DR7 at z = 0.04.}
    \end{tablenotes}
\end{table}
\clearpage

\newpage


\label{lastpage}
\end{document}